\documentclass[onecolumn,12pt,journal,draftclsnofoot]{IEEEtran}
\IEEEoverridecommandlockouts

\usepackage{cite}
\ifCLASSINFOpdf
\usepackage[pdftex]{graphicx}
\else
\fi

\usepackage[cmex10]{amsmath}
\usepackage{amssymb} 
\usepackage{amsfonts} 
\usepackage{blkarray} 
\usepackage{xcolor}  

\usepackage{keyval}

\usepackage{subcaption} 
\usepackage{fixltx2e}

\usepackage{flushend}

\def\A{{\mathbf A}}

\def\C{{\mathbf C}}
\def\D{{\mathbf D}}

\def\F{{\mathbf F}}

\def\H{{\mathbf H}}
\def\I{{\mathbf I}}

\def\P{{\mathbf P}}
\def\Q{{\mathbf Q}}
\def\R{{\mathbf R}}
\def\S{{\mathbf S}}
\def\T{{\mathbf T}}

\def\W{{\mathbf W}}
\def\X{{\mathbf X}}
\def\Y{{\mathbf Y}}
\def\Z{{\mathbf Z}}

\def\0{{\mathbf 0}}
\def\1{{\mathbf 1}}

\def\r{{\mathbf r}}

\def\w{{\mathbf w}}
\def\x{{\mathbf x}}

\DeclareMathOperator{\trace}{tr}

\DeclareMathOperator{\sinc}{sinc}

\newcommand{\openone}{\leavevmode\hbox{\small1\normalsize\kern-.33em1}}

\newcommand{\snr}{{\rm snr}}

\newcommand{\ex}[1]{\mathbb{E}\left[ {#1}\right]}
\newcommand{\LG}{{\rm LG}}

\newtheorem{lemma}{Lemma}

\newtheorem{theorem}{Theorem}

\begin{document}
%
\title{Partial-Duplex Amplify-and-Forward Relaying: Spectral Efficiency Analysis under Self-Interference }

\author{Roberto L\'opez-Valcarce,~\IEEEmembership{Member,~IEEE}, Carlos Mosquera,~\IEEEmembership{Senior~Member,~IEEE} \thanks{
Carlos Mosquera and Roberto L\'opez-Valcarce are with the Signal Theory and Communications Department, University of Vigo, Galicia, Spain. (e-mail: \{mosquera, valcarce\}@gts.uvigo.es).
This work was partially funded by the Agencia Estatal de Investigaci\'on (Spain) and the European Regional Development Fund (ERDF) through the projects MYRADA (grant TEC2016-75103-C2-2-R), WINTER (grant TEC2016-76409-C2-2-R) and COMONSENS (grant TEC2015-69648-REDC). Also funded by the Xunta de Galicia (Agrupaci\'on Estrat\'exica Consolidada de Galicia accreditation 2016-2019; Red Tem\'atica REdTEIC 2017-2018) and the European Union (European Regional Development Fund - ERDF).
Some preliminary results of this paper were presented in IEEE SPAWC 2016 \cite{MosqueraValcarce2016}.
}}

\maketitle

\begin{abstract}
We propose a novel mode of operation for Amplify-and-Forward relays in which the spectra of the relay input and output signals partially overlap. This {\em partial-duplex} relaying mode encompasses half-duplex and full-duplex as particular cases.  By viewing the partial-duplex relay as a bandwidth-preserving Linear Periodic Time-Varying system, an analysis of the spectral efficiency in the presence of self-interference is developed. In contrast with previous works, self-interference is regarded as a useful information-bearing component rather than simply assimilated to noise. This approach reveals that previous results regarding the impact of self-interference on (full-duplex) relay performance are overly pessimistic. Based on a frequency-domain interpretation of the effect of self-interference, a number of suboptimal decoding architectures at the destination node are also discussed. It is found that the partial-duplex relaying mode may provide an attractive tradeoff between spectral efficiency and receiver complexity.

\end{abstract}

\IEEEpeerreviewmaketitle

\section{Introduction}

Relay-assisted communication is a widespread technique to extend the coverage and improve the reliability of wireless networks \cite{Wornell2004,Soldani2008}. Depending on how the received signal is processed by the relay node, a number of relaying schemes can be identified. Among these, Amplify-and-Forward (A\&F), in which the relay just amplifies the received signal and then forwards it to the destination, emerges as a highly flexible technology, which is transparent to the particular modulation type of the retransmitted signal and has low implementation complexity \cite{Berger2009,Leonardo2015}. Traditionally, A\&F relays operate in Half-Duplex (HD) mode, meaning that  they transmit and receive either at different times, or over sufficiently separated frequency bands. This is because of the widespread belief that simultaneously transmitting and receiving on the same band would result in strong self-interference many tens of dB above the signal of interest, thus overwhelming the receiver. This Full-Duplex (FD) mode, however, is of great interest for next-generation wireless systems due to its potential to improve spectral efficiency by avoiding the use of additional time or frequency resources \cite{Sabharwal2014,Liu2015,Hanzo2016,Wichman2017}. Motivated by this, significant effort has been dedicated to the study of self-interference cancellation technologies \cite{Riihonen2011b,Duarte2012,Bharadia2013,Heino2015}, with results suggesting that operation in FD mode might be feasible. In fact, FD A\&F relaying is already found in certain practical settings such as on-frequency repeaters for broadcasting applications \cite{Shibuya2006,Choi2011,LopezValcarce2012}. Nevertheless, given the high levels of self-interference present in practice, some amount of residual self-interference is to be expected in most scenarios. For example, in a practical deployment of an FD A\&F relay, the designer may choose not to incorporate active cancellation methods in order to keep down complexity and cost, as long as sufficient mitigation of self-interference is provided by passive means (e.g. antenna placement and radiation pattern optimization) to avoid saturation of the receive analog frontend \cite{Everett2014}. Even when active suppression is applied, perfect cancellation is generally not possible \cite{Hanzo2016,Wichman2017}. Thus, analyzing the impact of self-interference in the performance of FD systems in general, and in FD A\&F relay networks in particular, has significant interest.

A number of such analyses, under different assumptions, can be found in the literature. In many of these, the residual self-interference is modeled as (usually Gaussian) noise, statistically independent of the retransmitted signal, and whose power depends in some way on the power of the signal transmitted by the relay~\cite{Shende2013,Leonardo2014a,Leonardo2014b,Zlatanov2017,Quek2017}. The Gaussian assumption is usually justified by invoking the Central Limit Theorem, given the variety of sources of imperfection in the cancellation process; whereas the independence assumption may be motivated by assuming a sufficiently large processing delay which effectively decorrelates the relay transmit signal with the simultaneously received signal \cite{Riihonen2011a}. 
In fact, this relay processing delay lies at the core of the problem because, as shown in \cite{Riihonen2009b,Kang2009}, when the processing delay is negligible (in the sense that the delay-bandwidth product is much smaller than one) self-interference ceases to be harmful as its effect can be assimilated to a mere scaling.

All the aforementioned works hinge on the assumption that self-interference can be regarded as noise. This is rather pessimistic, because the self-interference waveform contains useful information about the signal transmitted by the source. To the best of our knowledge, the impact on spectral efficiency  for an FD A\&F relay, assuming that the receiver at the destination is able to exploit self-interference in the decoding process, has not been studied yet. Our analysis shows that such approach results in a much more graceful performance degradation  compared to the standard procedure of treating self-interference as noise.

Specifically, in this work we study the performance of a single-input single-output (SISO) A\&F relay under a novel operation mode termed {\em Partial Duplexing} (PD), which encompasses HD and FD as particular cases. In PD mode, the relay transmits and receives simultaneously, placing the transmitted signal in a frequency band that {\em partially} overlaps with that of the incoming signal. In this way, HD and FD are obtained as particular instances of PD with zero and 100\% overlap, respectively. We evaluate the achievable rate as a function of the spectrum overlap factor, in the presence of self-interference, and for a variety of decoding strategies at the destination. We must note that PD operation is fundamentally different from previous hybrid HD/FD approaches \cite{Yamamoto2011,Riihonen2011a} that opportunistically switch between HD and FD modes depending on the quality of the different links in the relay network; additionally, channel state information (CSI) is not needed at the relay in PD mode, so that relay operation remains simple. The burden of the decoding process is placed at the destination node and increases with the spectrum overlap factor: therefore, judicious selection of this factor allows to trade off complexity and performance.

Since our study focuses on the A\&F relay itself, the source-to-relay, relay-to-destination, and self-interference channels are assumed frequency flat, with uniform power allocation over frequency at the source, and non-negligible processing delay at the relay. The source-to-destination link is assumed absent, and additive white Gaussian noise (AWGN) will be present only at the destination, as we assume that the signal-to-noise ratio (SNR) at the input of the relay is sufficiently large as a result of careful relay deployment. 
A key technical step in the development will be the consideration of the PD relay as a special type of Linear Periodically Time-Varying (LPTV) system \cite{Ericson1981,Mitra1988,Vanassche2002}, for which the bandwidth of the input signal is preserved at its output.

The paper is organized as follows. After introducing the PD relay operation in Sec. \ref{sec:PDrelay}, its spectral efficiency is analyzed in Secs. \ref{sec:TDapproach} and \ref{sec:FDapproach} using a time-domain approach and a frequency-domain approximation, respectively. A number of suboptimal receiver architectures with different levels of complexity are discussed in Sec.~\ref{sec:receiver}. In Sec.~\ref{sec:HDvsFD} a comparison between HD and FD modes is provided. Numerical results are given in Sec.~\ref{sec:simulations}, and conclusions are drawn in Sec.~\ref{sec:conclusions}.\\

\section{Partial Duplex Relay}\label{sec:PDrelay}
\label{sec:PDrelay}

\begin{figure*}
\centering
\includegraphics[width=0.9\textwidth]{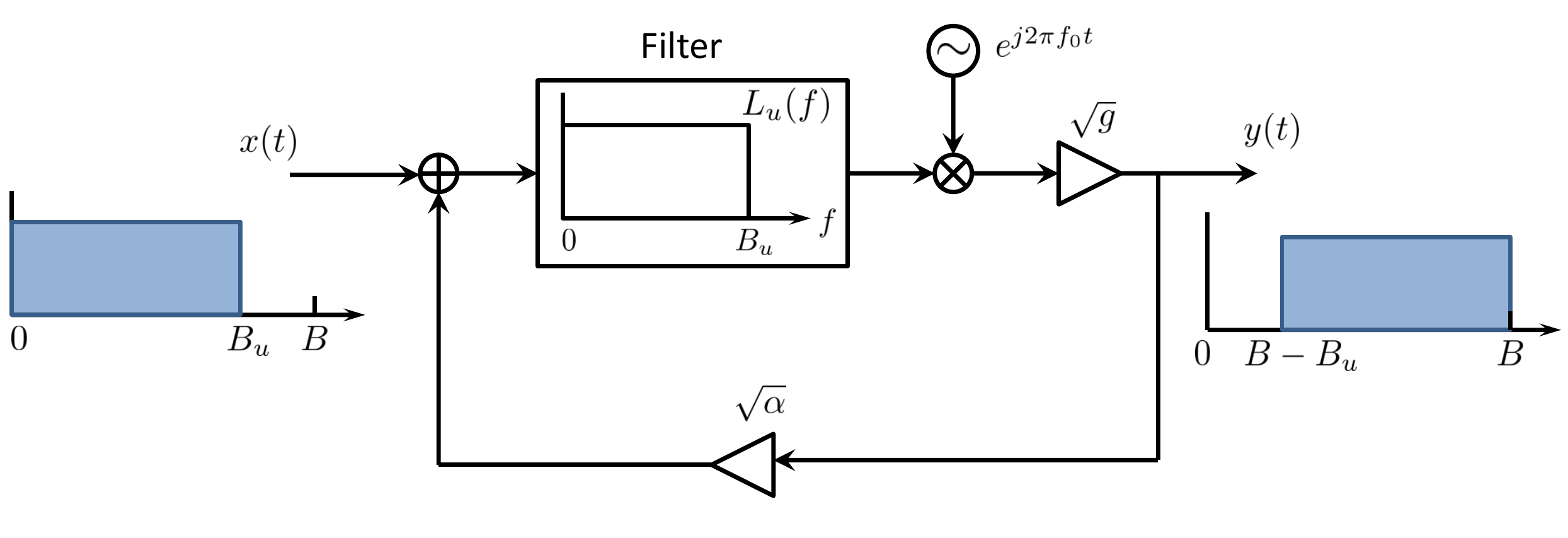}
\caption{Baseband-equivalent description of the Partial Duplex A\&F relay. In practice, the receive and transmit front-ends incorporate a down-conversion and up-conversion stage, respectively. The difference between the corresponding oscillator frequencies is given by $f_0$.}
\label{fig:relay}
\end{figure*}

\subsection{System model}
Fig. \ref{fig:relay} shows the operation of the proposed PD A\&F relay. The source transmits a signal $x(t)$, with bandwidth $B_u$ and power $\bar P_x$. Upon reaching the relay input, this signal is filtered, frequency-shifted by $f_0 = B-B_u$, and amplified.
As long as the passband of the output signal overlaps that of the input, i.e., $f_0<B_u$, self-interference will be present due to coupling from the relay output to its input. The power gain of the self-interference path is denoted by $\alpha$. The purpose of the relay filter is to eliminate as much self-interference as possible; we assume an ideal frequency response given by 
\begin{equation}
L_u(f) = \left\{
\begin{array}{cc}
e^{-j (2\pi f t_0+\theta_0)}, & 0 \leq f \leq B_u \\
0, & \mbox{otherwise,}
\end{array}
\right.
\label{eq:Lu}
\end{equation}
where $\theta_0$ is a constant phase shift and $t_0$ is the group delay.  In practice, the time delay of the self-interference path is generally much smaller than the group delay of the analog filters in the relay frontends (i.e., $t_0$), and thus  it is neglected.

After filtering, the signal is frequency-shifted by multiplication with $e^{j(2\pi f_0t + \theta_{\rm OL})}$; since any constant phase offset $\theta_{\rm OL}$ in the local oscillator can be absorbed in the parameter $\theta_0$ in \eqref{eq:Lu}, we will assume without loss of generality that $\theta_{\rm OL}=0$.
Finally, the signal is amplified with power gain $g$ to yield output power $\bar P_y$; the retransmitted signal is denoted by $y(t)$.
At the destination, the output signal $y(t)$ is corrupted by additive white Gaussian noise. 
It is assumed that the source-to-relay and relay-to-destination links are frequency flat, and that the direct link from the source to the destination is absent.

Given the total system bandwidth $B$, the design parameter is the bandwidth allocated to the input and output signals, $B_u = B - f_0$. This can be expressed in terms of the ratio
\begin{equation}
\rho \triangleq \frac{B_u}{B} \, \in \, \left[\tfrac{1}{2}, 1\right].
\end{equation}
Under the configuration in Fig. \ref{fig:relay}, the HD mode corresponds to $B_u = B/2$ (i.e., $f_0=B/2$, or $\rho=\tfrac{1}{2}$), with non-overlapping input and ouptut spectra; 
whereas the FD mode is recovered for $B_u = B$ (i.e., $f_0=0$, or $\rho=1$), with input and output passbands completely overlapping. For $\tfrac{1}{2} < \rho < 1$, the operational mode is termed {\em Partial Duplex} (PD). 

The operation of the PD relay can be written in the frequency domain as 
\begin{equation}\label{eq:ioFourier}
Y(f) = \sqrt{g} X(f-f_0) L_u(f-f_0) + \sqrt{\alpha g} Y(f-f_0) L_u(f-f_0),
\end{equation}
a recursive relation which can be unfolded to yield 
\begin{multline}\label{eq:iorelation}
Y(f) = \sqrt{g} X(f-f_0) L_u(f-f_0) + \\
\overbrace{\sqrt{g}\sum_{k=1}^{K} (\sqrt{\alpha g})^k X(f-(k+1) f_0) \Pi_{m=1}^{k+1} L_u(f-mf_0)}^{\text{self-interference}}.
\end{multline}
The number of terms contributing to the self-intererence sum  is  finite except for $B_u = B$ (FD case), and it is given by 
\begin{equation}\label{eq:Kmax}
K \triangleq \left\lceil{\frac{B_u}{B-B_u}} \right \rceil - 1 \,=\,  \left\lceil{\frac{\rho}{1-\rho}} \right \rceil - 1,
\end{equation} 
where $\left\lceil{\cdot}\right\rceil$ denotes the ceil function. This is readily obtained from the fact that the filter $L_u(f)$ has a bandwidth of $B_u$, whereas each time the self-interference signal loops through the coupling path it undergoes a frequency shift of $f_0=B-B_u$ (see Fig.~\ref{fig:relay}).

We assume that the relay uses automatic gain control (AGC), as customary in practical repeaters. Thus, the power gain $g$ is not set beforehand, but rather it is adjusted to deliver the nominal output power $\bar P_y$ even in the presence of coupling from output to input. In the particular case without self-interference ($\alpha=0$), the relay gain becomes simply $g = \bar P_y/\bar P_x$.

As performance metric we consider the achievable rate from the source to the final destination through the PD relay, in the absence of a direct link. 
We will assume that the relative contributions of all noise sources are much less significant than that of the final receiver noise, since proper operation of repeaters usually require a good source-relay link.  
Assuming white noise at the destination with power spectral density $N_0$, we define the {\em reference signal-to-noise ratio} and the {\em loop gain} respectively as
\begin{equation}\label{eq:snrlg}
\snr \triangleq   \frac{\bar P_y}{N_0 B}, \qquad \LG \triangleq \frac{\alpha \bar P_y}{\bar P_x}.
\end{equation}
Note that $\frac{1}{\LG}$ can be interpreted as the signal-to-self-interference ratio at the relay input.

It is illustrative to analyze the power budget of the AGC-equipped relay in the FD case ($B_u=B$, or $\rho=1$), for which the input-output relation \eqref{eq:iorelation} becomes, after letting $f_0\to 0$ and $K\to\infty$,
\begin{equation}\label{eq:iorelationFD}
Y(f) = \frac{\sqrt{g} e^{-j(2\pi ft_0+\theta_0)}}{1-\sqrt{\alpha g}e^{-j(2\pi ft_0+\theta_0)}} X(f) = H(f)X(f),
\end{equation}
showing that the FD relay is linear time-invariant (LTI) with transfer function $H(f)$. 
From \eqref{eq:iorelation}, if we regard the term $\sqrt{g} e^{-j(2\pi ft_0+\theta_0)} X(f)$ as the useful signal component (with power $g \bar P_x$) at the relay output, then the spectrum of the remaining terms (self-interference) is given by 
\begin{eqnarray}
Y(f) - \sqrt{g} e^{-j(2\pi ft_0+\theta_0)} X(f) &=& \left[H(f) - \sqrt{g} e^{-j(2\pi ft_0+\theta_0)}\right]X(f) \\
&=& \sqrt{\alpha g} e^{-j(2\pi ft_0+\theta_0)} H(f) X(f),
\end{eqnarray}
from which the power of the self-interference is seen to be given by $\alpha g \bar P_y$. The signal-to-self-interference ratio at the relay output is therefore $\frac{g\bar P_x}{\alpha g \bar P_y} = \frac{1}{\LG}$, i.e., the same as that at its input. If self-interference is assimilated to noise, then the performance of the FD relay can be expected to degrade fast as the loop gain increases, leading to overly pessimistic results.


\subsection{PD relay as an LPTV system}
For the subsequent  analysis, it is important to note  that for $B/2 < B_u < B$ (i.e., $\frac{1}{2}<\rho<1$) the PD relay is not LTI, 
but rather Linear Periodically Time-Varying (LPTV). 
The input-output frequency relationship of a generic LPTV system with period $T_0$ \cite{Vanassche2002,Mosquera2002} can be written as 
\begin{equation}\label{eq:LPTV}
Y(f) = \sum_{k=-\infty}^\infty H_k(f) X\left(f-\frac{k}{T_0}\right)
\end{equation}
for some transfer functions $\{H_k(f)\}$. By comparing \eqref{eq:LPTV} with \eqref{eq:iorelation}, and defining
\begin{equation}\label{eq:Lk}
L_k(f) \triangleq \prod_{m=1}^k L_u(f-mf_0), \quad k=1,\ldots,K+1,
\end{equation}
it is clear that the PD relay is LPTV with period $T_0=1/f_0$ and 
\begin{equation}\label{eq:Hk}
H_k(f) = \left\{\begin{array}{cc} \sqrt{g}\left(\sqrt{\alpha g}\right)^{k-1} L_k(f), & k=1,\ldots,K+1, \\ 0, & \mbox{otherwise.} \end{array}\right.
\end{equation}
Note that the support of $H_k(f)$ is within the interval $[B-B_u, B]$.
Thus, assuming an ideal filter $L_u(f)$, the PD relay belongs to the particular class of {\em bandwidth-preserving} LPTV systems, since the size of the spectral region with non-zero frequency content is the same for both input and output signals, namely $B_u$.
With non-ideal filters some out-of-band content will appear, resulting in a spectral efficiency loss.


\section{Spectral Efficiency of PD Relay: Time-Domain Approach}
\label{sec:TDapproach}
Somewhat surprisingly, the information-theoretic analysis of LPTV channels has not been directly addressed in the literature until recently.  The following derivation follows the steps of the time-domain approach from \cite{Dabora2015}, which dealt with Power Line Communication (PLC) channels, and was based on the assimilation of the SISO LPTV
channel to a multiple-input multiple-output (MIMO) LTI system. This approach can be traced back to \cite{Wyner1974}, which obtained the capacity of the multivariate Gaussian channel with memory by formulating the input-output relationship as a memoryless MIMO channel (although the term "MIMO" was not used at that time).

\begin{figure*}
\centering
\includegraphics[width=0.9\textwidth]{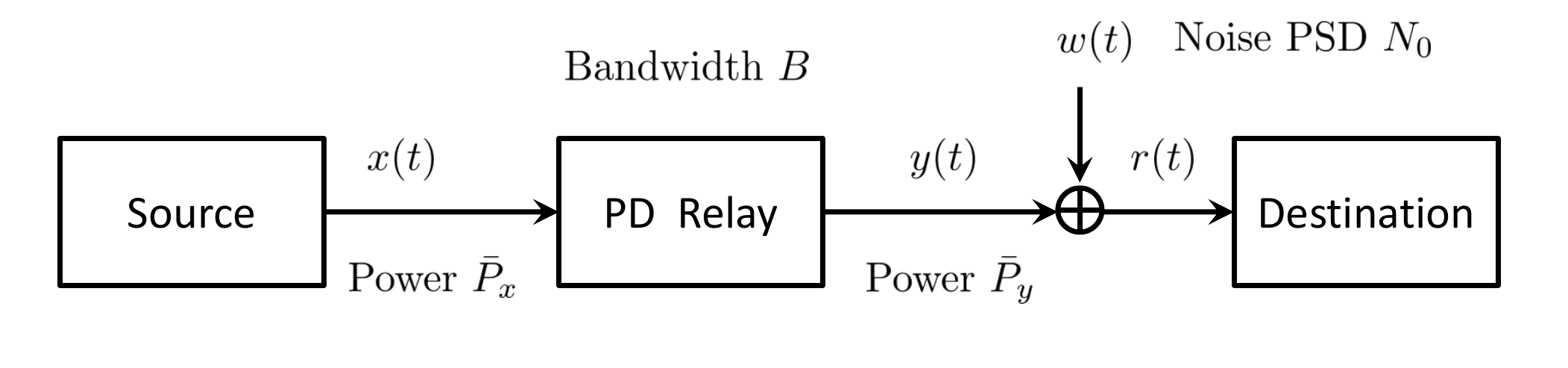}
\caption{The source communicates with the destination through a SISO relay with output power $\bar P_y$. Part of the retransmitted signal loops back to the relay input, resulting in self-interference. }
\label{fig:relaychain}
\end{figure*}

The overall system including the source, PD relay, and destination is shown in Fig.~\ref{fig:relaychain}. The received signal at the destination can be written as 
\begin{equation}\label{eq:received}
r(t) = y(t) + w(t)= \sqrt{g} \int_{-\infty}^\infty p(t,\tau) x(t-\tau) d\tau + w(t),
\end{equation}
which is the time-domain counterpart of  \eqref{eq:iorelation}, with the addition of the noise $w(t)$. 
An ideal filter with passband $[0,B]$ is assumed at the receiver, so that the psd of the noise $w(t)$ is $N_0$ for $0\leq f\leq B$ and zero otherwise.
In terms of the impulse responses $l_k(t)=\int_{-\infty}^\infty L_k(f) e^{j2\pi ft}df$, $k=1,\ldots,K+1$, the response $p(t,\tau)$ is given by 
\begin{equation}
p(t,\tau) = \sum_{k=1}^{K+1} \left(\sqrt{g \alpha}\right)^{k-1} l_k(\tau) e^{j 2\pi k f_0(t-\tau)}.
\end{equation}
Note that the input-output relation \eqref{eq:received} does correspond to an LPTV system, since $p(t,\tau) = p(t+T_0,\tau)$ with $T_0=1/f_0$. In discrete-time form, if the sampling rate is $1/T$, we have 
\begin{equation}\label{eq:sampled}
y(nT) = \sqrt{g} \sum_{m=-\infty}^\infty p(nT,mT) x((n-m)T)
\end{equation} 
with the time-varying discrete-time impulse response 
\begin{equation}\label{eq:pwithLk}
p(nT,mT) = \sum_{k=1}^{K+1} \left(\sqrt{g \alpha}\right)^{k-1} l_k(mT)
e^{j 2\pi k f_0T(n-m)}.
\end{equation}
Note that, upon choosing $T \leq 1/B$, the Nyquist criterion is satisfied for all bandwidths under consideration in Fig. \ref{fig:relay}. In addition, the sampled system \eqref{eq:sampled} will be LPTV with period $r \in \mathbb{N}$ provided that $T = T_0/r$. These two conditions will be simultaneously satisfied if $r\geq \left\lceil\frac{B}{B-B_u} \right\rceil = K+2$. In this section we will assume that $\frac{B}{B-B_u}=N_{ch}$ is an integer (or equivalently, that $\rho = \frac{N_{ch}-1}{N_{ch}}$ for some integer $N_{ch}$, or $B_u = \left(1-\frac{1}{N_{ch}}\right)B$~), and choose $r=N_{ch}$ so that $T=T_0/N_{ch} = 1/B$. 
In this way, it can be readily checked that the noise samples $w(nT)$, with variance $BN_0$, will be uncorrelated, which simplifies the analysis. In Sec. \ref{sec:FDapproach} a frequency-domain approximation will be presented, which allows to extend the results to the case in which $\frac{B}{B-B_u}$ is not an integer.

Thus, with $f_0T = 1/N_{ch}$, \eqref{eq:pwithLk} becomes
\begin{eqnarray}\label{eq:psimple}
  p(nT,mT) &=& 
 \sum_{k=1}^{K+1} (\sqrt{g \alpha})^{k-1}  l_k(mT) e^{-j \frac{2\pi}{N_{ch}}km} e^{j \frac{2\pi }{N_{ch}}kn} \nonumber\\
&\triangleq& p_n(mT), \qquad n=0,1,\ldots, N_{ch}-1.
\end{eqnarray}
Let us define the delay in samples as $\ell \triangleq t_0/T$.
For the ideal filter response (\ref{eq:Lu}), it can be checked that the term $l_k(mT) e^{-j  \frac{2\pi }{N_{ch}}km}$ in \eqref{eq:pwithLk} reads as 
\begin{equation}\label{eq:lke}
l_k(mT) e^{-j \frac{2\pi }{N_{ch}}km} = e^{-jk\theta_0}
e^{j\pi\left(1 - \frac{k}{N_{ch}}\right)\left(m-k\ell\right)} \left(1 - \frac{k}{N_{ch}}\right)
\sinc \left[ \left(1 - \frac{k}{N_{ch}}\right)\left(m-k\ell\right)\right].
\end{equation}
If the delay-bandwidth product of the relay filter is large, i.e., if $Bt_0 = \ell \gg 1$, then \eqref{eq:lke} is approximately zero outside the interval $0 \leq m \leq 2 k \ell$, and it follows that $p_n(mT) \approx 0$ outside the interval $0 \leq m \leq \ell_p$, with $\ell_p \triangleq 2(K+1)\ell$,  for all $n=0,1,\ldots, N_{ch}-1$.

To find the capacity of the PD relay channel, the original scalar model is first transformed into a vector model. Let $M$ be the size of the input block, defined as
\begin{equation}\label{eq:inputblock}
\x[n] \triangleq \left[\begin{array}{cccc} x(nMT) & x((nM+1)T) & \cdots & x((nM+M-1)T) \end{array} \right]^T.
\end{equation}
Similarly, we define the output block and noise vector respectively as
\begin{eqnarray}
\r[n] &\triangleq & \left[\begin{array}{cccc} r((nM+\ell_p)T) & r((nM+\ell_p+1)T) & \cdots & r((nM+M-1)T) \end{array} \right]^T, \\
\w[n] &\triangleq & \left[\begin{array}{cccc} w((nM+\ell_p)T) & w((nM+\ell_p+1)T) & \cdots & w((nM+M-1)T) \end{array} \right]^T,
\end{eqnarray}
both having size $M-\ell_p$. Then, the input-output relationship can be expressed in matrix form as
\begin{equation}\label{eq:iotimerelation}
\r[n]= \sqrt{g} \P \x[n] + \w[n],
\end{equation}
with the $(M-\ell_p)\times M$ channel matrix $\P$ given by 
\begin{equation}\label{eq:P}
\P \triangleq \left(
\begin{array}{ccccc}
p_{\ell_p}(\ell_pT) & \hdots & p_{\ell_p}(0) & \hdots & 0 \\
\vdots             & \ddots &            & \ddots & \vdots \\
0                 & \hdots & p_{M-1}(\ell_pT) & \hdots & p_{M-1}(0)
\end{array}
\right),
\end{equation}
where it is implicitly assumed that $p_n(mT)$ is $N_{ch}$-periodic in $n$. The block size $M$ is chosen as an integer multiple of $N_{ch}$,
so that the input block comprises an integer number of periods.
Note that the size of the output block, $M-\ell_p$, is smaller than that of the input block, $M$. 
Nevertheless, the impact on the capacity derivation decreases as the block size $M$ grows, and the true capacity $C$ can be obtained as the asymptotic value  $\lim_{M \rightarrow \infty} C_M$ \cite{Dabora2015}, with $C_M$ denoting the achievable rate of the truncated MIMO model with channel matrix $\P$ in (\ref{eq:P}) 
for a given input covariance 
$\C_x \triangleq \ex {\x[n] \x^H[n]}$. Since $E\{\w[n]\w^H[n]\}= BN_0 \I$,
\begin{equation} \label{eq:bpcu}
C_M  = \frac{1}{(M-\ell_p)T}\log_2 \det \left(\I + \frac{g}{BN_0} \P \C_x\P^H    \right) \qquad [\mbox{bps}].
\end{equation}
Therefore, with $BT=1$, the corresponding spectral efficiency is
\begin{equation}\label{eq:Clptv}
\frac{C}{B} = \lim_{M \rightarrow \infty} \frac{1}{M-\ell_p}\log_2 \det \left(\I + \frac{g}{BN_0} \P \C_x\P^H    \right) \qquad [\mbox{bps/Hz}].
\end{equation}
Since we are considering frequency-nonselective channels, the source is assumed to transmit with flat power spectral density in the occupied bandwidth. Then $[\C_x]_{k,l} = C_x((k-l)T)$, with
\begin{equation}
C_x(\tau) \triangleq \bar P_x \, \sinc\left(B_u \tau\right) e^{j 2\pi  \frac{B_u}{2} \tau},
\end{equation}
which for $B_u= \left(1-\frac{1}{N_{ch}}\right)B$ and  $T=\frac{1}{B}$ yields
\begin{equation}
C_x(mT) = \bar P_x \, \sinc\left[\left(1-\frac{1}{N_{ch}}\right)m\right] e^{j \pi \left(1-\frac{1}{N_{ch}}\right)m}.
\end{equation}
In order to set the relay gain $g$, note that the relay output power must equal $\bar P_y$, i.e.,
\begin{equation}\label{eq:agc}
g \cdot \frac{\trace\{\P \C_x \P^H \}}{M-\ell_p} = \bar P_y.
\end{equation}
The entries of $\P$ are functions of $g$, and thus \eqref{eq:agc} is nonlinear in $g$ and has to be solved by numerical means. An alternative approach to obtain $g$ will be presented in Sec.~\ref{sec:FDapproach}. 

\begin{figure*}
\centering
\includegraphics[width=0.7\textwidth]{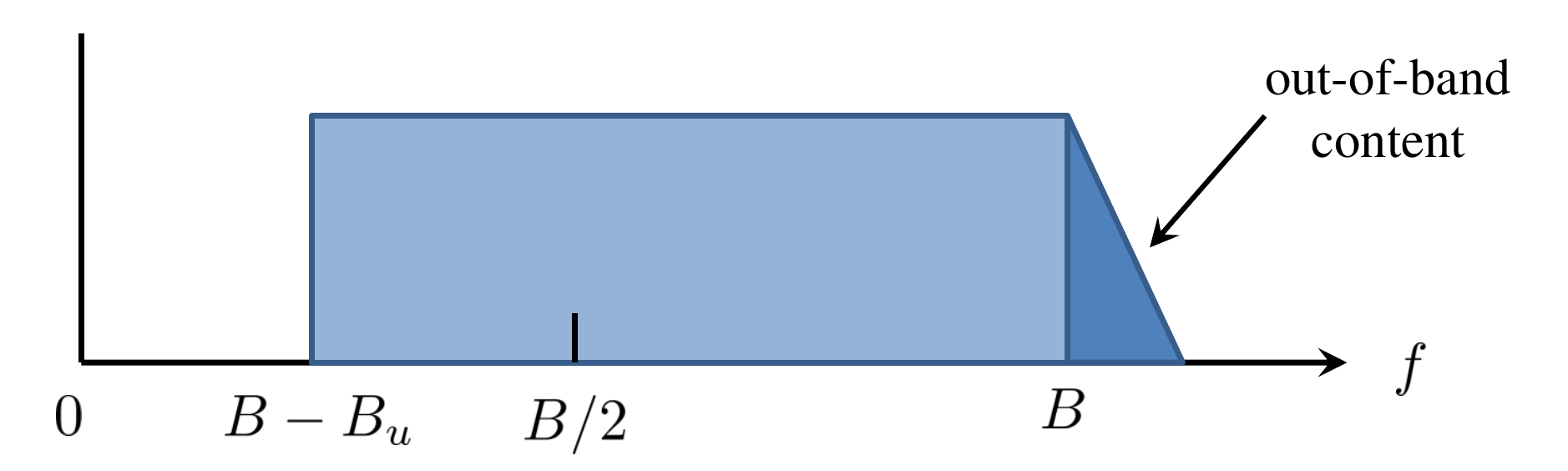}
\caption{A non-ideal relay filter results in some residual out-of-band content.}
\label{fig:outofband}
\end{figure*}

As a final remark, we mention that, with a non-ideal relay filter $L_u(f)$, some out-of-band content will be present at the relay output as shown in Fig.~\ref{fig:outofband}. 
The bulk of the above analysis should still hold, although the spectral efficiency in \eqref{eq:Clptv} will suffer a slight degradation due to the additional bandwidth taken by the residual content. 

\section{Spectral Efficiency of PD Relay: Frequency-Domain Approach}
\label{sec:FDapproach}
We present now an alternative approach to computing the capacity of the PD relay, based on the frequency-domain input-output relation \eqref{eq:LPTV}-\eqref{eq:Hk}, and following standard arguments for frequency-selective LTI channels \cite{tse2005}. The available bandwidth $B$ is sliced into a total of $L$ subcarriers, and the source transmits by using only $N$ of them, with $N<L$, while leaving the remaining $P\triangleq L-N$ subcarriers unused, such that $\frac{N}{L} = \frac{B_u}{B}=\rho$. Thus, the intercarrier spacing is $\Delta f=\frac{B}{L}=\frac{B_u}{N}$, and the frequency offset in Fig.~\ref{fig:relay} equals $f_0 = B-B_u = P \Delta f$.  
Transmission is block-based, with blocks of length $L$ to which a cyclic prefix of length $\ell_p$ is added. The overhead due to the cyclic prefix can be made arbitrarily small as $L\to\infty$.

For LTI channels, this multicarrier approach results in the familiar decoupling of the channel into a set of $L$ independent parallel subchannels with no intercarrier interference (ICI), and with the gain of each subchannel given by the transfer function of the channel at the corresponding frequency bin. However, for the PD relay with self-interference, ICI will be present due to the fact that the received signal spectrum $Y(f)$ is the superposition of a number of scaled and frequency-shifted replicas $H_k(f)X(f-kf_0)$ of the original spectrum $X(f)$ as seen in \eqref{eq:LPTV}. 

Again, let us assume a sampling rate $T=1/B$, so that the noise samples at the output of the receive filter (with passband $[0,B]$) are uncorrelated with power $BN_0$. If we let $\R[q]$, $\X[q]$ and $\W[q]$ respectively denote the $L\times 1$ vectors given by the Discrete Fourier Transform (DFT) of the $q$th output, input and noise blocks (of length $L$, or duration $LT$ seconds), then the input-output relation from $\X[q]$ to $\R[q]$ can be well approximated, for sufficiently large $L$, as 
\begin{equation}\label{eq:Frelation}
\R[q] = \H \X[q] + \W[q], 
\end{equation}
where $\H \in \mathbb{C}^{L\times L}$ comprises the ICI coefficients:
\begin{equation}
[\H]_{n,m} = \left\{\begin{array}{cc} H_k(n\Delta f), & \mbox{if~} m=n-kP \mbox{~with~} 1\leq k \leq K+1,\\
			0, & \mbox{otherwise,} \end{array}\right.
\end{equation}
with $0\leq n,m\leq L-1$, and $K$ given by \eqref{eq:Kmax} or, equivalently, $K = \left\lceil \frac{N}{P}\right\rceil -1$.
 In view of \eqref{eq:Hk}, $\H$ is seen to be lower triangular; moreover, its first $P$ rows and last $P$ columns are zero. Since the source does not use the last $P$ subcarriers, i.e., the last $P$ entries of $\X[q]$ are zero, it follows that $\H\X[q]$ can be written as:
\begin{equation}\label{eq:HX}
 \H\X[q] = \left(\begin{array}{c|c}
 \0_{P\times N}  & \0_{P\times P} \\
\hline
\sqrt{g}\T_{N}\D_N  & \0_{N \times P} 
\end{array}
\right)
\left(\begin{array}{c} \bar\X[q] \\ \0_{P}\end{array}\right)
= \left(\begin{array}{c} \0_{P} \\ \bar\Y[q] \end{array}\right),
\end{equation}
where $\bar\X[q]$ and $\bar\Y[q]$ are $N\times 1$, $\T_N \in\mathbb{C}^{N\times N}$ is lower triangular, and $\D_N \in\mathbb{C}^{N\times N}$ is diagonal. The entries of these matrices are as follows: letting $\phi_0 \triangleq 2\pi \Delta f t_0$, one has
\begin{equation}
[\D_N]_{n,n} = e^{-j(n\phi_0+\theta_0)},\qquad 0 \leq n \leq N-1.
\end{equation}
The entries $[\T_N]_{n,m}$, $0\leq n,m\leq N-1$, are zero except when $m=n-kP$ for some $k \in\{ 0,1,\ldots,K\}$, in which case one has
\begin{equation}\label{eq:Tentries}
[\T_N]_{n,n-kP} = \left(\sqrt{\alpha g}\right)^k e^{-jk\theta_0} e^{-j\left(nk-P\frac{k(k-1)}{2}\right)\phi_0}.
\end{equation}
Note in particular that $\D_N\D_N^H = \I_N$ and $\T_N$ has ones on its diagonal.

From \eqref{eq:Frelation} and \eqref{eq:HX} it follows that
\begin{equation}\label{eq:Frelation2}
\bar\R[q] = \sqrt{g} \T_N \D_N \bar\X[q] + \bar\W[q], 
\end{equation}
where $\bar\R[q]$ and $\bar\W[q]$ comprise the last $N$ entries of $\R[q]$ and $\W[q]$, respectively. The noise vector $\bar\W[q]$ is zero-mean Gaussian with covariance matrix $BN_0 \I_N$, whereas that of $\bar\X[q]$ is $\frac{L}{N}\bar P_x\I_N$ for constant power allocation across $f\in [0, B_u]$ at the source\footnote{This follows from the fact that the source transmit power is $\bar P_x = \frac{1}{L}\trace E\{\X[q]\X^H[q]\}$ and that the last $P=L-N$ entries of $\X[q]$ are zero.}.

The sum rate of all carriers is bounded by  the relay channel capacity, which from (\ref{eq:Frelation2}) gives the following achievable spectral efficiency:  
\begin{eqnarray}
\frac{C}{B} &=& \frac{1}{L}   \log_2 \det \left(\I_N + g\frac{L}{N} \frac{\bar P_x}{B N_0}  \T_N \T_N^H    \right)
\label{eq:Cpfd} \\
&=& \frac{1}{L}   \log_2 \det \left(\I_N + \mu \T_N \T_N^H    \right), \label{eq:Cpfd2}
\end{eqnarray}
where we have introduced 
\begin{equation}\label{eq:mudef}
\mu \triangleq g\frac{L}{N}\frac{\bar P_x}{BN_0} = \alpha g \frac{\snr}{\rho \LG},
\end{equation}
which is a scaled version of the reference SNR \eqref{eq:snrlg}, and with $\LG$ the loop gain, defined in \eqref{eq:snrlg}. The gain $g$ can be obtained from the expression of the PD relay output power, given by 
\begin{eqnarray}
 \bar P_y &=& \frac{1}{L} \trace E\{\H\X[q]\X^H[q]\H^H\} \nonumber \\
 &=& \frac{g}{L} \trace E\{\T_N \bar \X[q]\bar \X^H[q] \T_N^H\}\nonumber \\
 &=& \frac{g}{N} \bar P_x \trace \{\T_N \T_N^H\}. \label{eq:Pyg}
\end{eqnarray}
Hence, $g = \frac{\bar P_y}{\bar P_x} \frac{N}{\trace \{\T_N \T_N^H\}}$. 
From the expression of $\T_N$, and using $\frac{P}{N} = \frac{1-\rho}{\rho}$, one readily finds that 
\begin{equation}\label{eq:traceTTH}
\frac{1}{N}\trace\{\T_N\T_N^H\} = \frac{1}{\rho}\sum_{k=0}^K(\rho-k(1-\rho))(\alpha g)^k.
\end{equation}
From \eqref{eq:Pyg} and \eqref{eq:traceTTH}, it follows that $\alpha g$ is a solution of the polynomial equation 
\begin{equation}\label{eq:AGCpoly}
\sum_{k=1}^{K+1}(1-k(1-\rho))(\alpha g)^k = \rho \LG.
\end{equation}
The following lemma establishes the uniqueness of the solution; see Appendix \ref{app:uniqueness} for the proof.
\begin{lemma}\label{lem:uniqueness} Assume $\alpha>0$. Then the polynomial equation \eqref{eq:AGCpoly} has a single solution satisfying $\alpha g >0$. In particular, if $\rho=1$ (FD case), the solution is $\alpha g = \frac{\LG}{1+\LG}$. 
If $\alpha=0$ (no self-interference), then $g=\frac{\bar P_y}{\bar P_x}$.
\end{lemma}

Now, in order to further develop \eqref{eq:Cpfd2}, let 
\begin{equation}
\Q_N \triangleq \left(\T_N\T_N^H\right)^{-1}, \label{eq:Qdef}
\end{equation}
with characteristic polynomial $q(\lambda) \triangleq \det(\Q_N - \lambda \I_N)$.
Then, the determinant in \eqref{eq:Cpfd2} can be written as
\begin{equation}\label{eq:detdet}
\det\left( \I_N + \mu\T_N\T_N^H \right) = \frac{\det\left( \Q_N + \mu\I_N \right)}{\det(\Q_N)} = \frac{q(-\mu)}{q(0)}.
\end{equation}
The usefulness of \eqref{eq:detdet} resides in the fact that $\Q_N$ has certain structure that we will exploit. To this end, the following lemma will be useful.
\begin{lemma}\label{lem:inverse}
The inverse of $\T_N$ is $\T_N^{-1} = \I_N - \sqrt{\alpha g}e^{-j\theta_0}\S_N$, where $\S_N$ is defined entrywise as
\begin{equation}\label{eq:Sentries}
[\S_N]_{n,n-P} =  e^{-jn\phi_0}, \quad n=P,P+1,\ldots, N-1;
\qquad [\S_N]_{n,m}=0, \quad \mbox{otherwise.}
\end{equation}
\end{lemma}
Lemma \ref{lem:inverse} is proved by using \eqref{eq:Tentries} and \eqref{eq:Sentries}  to directly verify that $\T_N(\I_N - \sqrt{\alpha g}e^{-j\theta_0}\S_N) = \I_N$. Using this result, one finds that
\begin{eqnarray}
\Q_N &=& (\I_N - \sqrt{\alpha g}e^{j\theta_0}\S_N^H)(\I_N - \sqrt{\alpha g}e^{-j\theta_0}\S_N) \\
&=& \I_N - \sqrt{\alpha g}e^{j\theta_0}\S_N^H - \sqrt{\alpha g}e^{-j\theta_0}\S_N + \alpha g \S_N^H\S_N.
\end{eqnarray}
Note now that the matrix $\S_N^H\S_N$ is diagonal, with the first $N-P$ diagonal elements equal to one and the last $P$ equal to zero.
Therefore, the entries of $\Q_N$ are zero except along the main diagonal and the $P$-th super- and sub-diagonals:
\begin{eqnarray}
\left[\Q_N\right]_{n,n} &=& \left\{\begin{array}{cc} 1+\alpha g, & 0 \leq n \leq N-P-1, \\ 1, & N-P \leq n \leq N-1, \end{array}\right. \label{eq:Qa} \\
\left[ \Q_{N}\right]_{n,n-P}  &=& -\sqrt{\alpha g} e^{-j(\theta_0 + n\phi_0)}, 
\quad \left[\Q_N\right]_{n-P,n} = \left[\Q_N\right]_{n,n-P}^*, \quad P\leq n \leq N-1, \label{eq:Qb}\\
\left[\Q_N\right]_{n,m}  &=& 0, \qquad \mbox{otherwise.} \label{eq:Qc}
\end{eqnarray}
The following result holds now; the proof hinges on the structure of $\Q_N$ as exposed by \eqref{eq:Qa}-\eqref{eq:Qc} and can be found in Appendix \ref{app:detQ}.
\begin{theorem} \label{th:detQ}
Let $q_0(\lambda)=1$, $q_1(\lambda) = 1-\lambda$ and
\begin{equation}\label{eq:recursionq}
q_{k}(\lambda) = (1+\alpha g - \lambda) q_{k-1}(\lambda) - \alpha g\, q_{k-2}(\lambda), \qquad k \geq 2.
\end{equation}
Then, with $K=\left\lceil \frac{N}{P}\right\rceil - 1$, the characteristic polynomial $q(\lambda) = \det(\Q_N - \lambda \I_N)$ is given by
\begin{equation}
q(\lambda) = \left[ q_{K}(\lambda) \right]^{(K+1)P-N} \left[ q_{K+1}(\lambda) \right]^{N-KP}.
\end{equation}
\end{theorem}

Note in particular that, from Theorem \ref{th:detQ}, the coefficients of the polynomial $q(\lambda)$ are functions of $\alpha g$ alone, and they do not depend on $\theta_0$ or $\phi_0$. Additionally, it follows by induction on $k$ that $q_k(0) = 1$ for all $k$, so that $q(0)=1$. 
Therefore, from \eqref{eq:Cpfd}-\eqref{eq:detdet}, the following expression for the spectral efficiency follows:
\begin{equation}\label{eq:speff1}
\frac{C}{B} = \frac{1}{L}\log_2 q(-\mu) = \frac{(K+1)P-N}{L}\log_2q_K(-\mu) +\frac{N-KP}{L}\log_2q_{K+1}(-\mu). 
\end{equation}
Let $\delta(\rho) \in [0,1)$ be the fractional part of $\frac{\rho}{1-\rho}$, i.e.,
\begin{equation}\label{eq:deltarho}
\delta(\rho) \triangleq \left\lceil \frac{\rho}{1-\rho}\right\rceil - \frac{\rho}{1-\rho}.
\end{equation}
Since $K = \left\lceil \frac{\rho}{1-\rho}\right\rceil -1$, one has $\frac{\rho}{1-\rho} = K+1-\delta(\rho)$, so that \eqref{eq:speff1}
can be compactly written as
\begin{equation}\label{eq:speff2}
\frac{C}{B} =   (1-\rho) \left[\,\delta(\rho) \log_2q_K(-\mu) + (1-\delta(\rho)) \log_2q_{K+1}(-\mu)\, \right]. 
\end{equation}
In the absence of self-interference ($\alpha=0$), one has $g=\bar P_y/\bar P_x$, $\mu = 1+\frac{\snr}{\rho}$ and $q_k(-\mu) = \left(1+\frac{\snr}{\rho}\right)^k$, $k\geq 0$, so that \eqref{eq:speff2} yields
\begin{equation}\label{eq:speff_no_si}
\frac{C}{B} =   \rho \log_2\left(1+\frac{\snr}{\rho}\right) \qquad \mbox{(no self-interference).}
\end{equation}
On the other hand, using the high-SNR approximation $1+\alpha g + \mu \approx \mu$ when computing $q_k(-\mu)$ in \eqref{eq:speff2} yields $q_k(-\mu) \approx \mu^k$, and therefore the asymptotic expression of the spectral efficiency \eqref{eq:speff2} for high SNR is
\begin{equation}\label{eq:Casymptotic}
\frac{C}{B} \approx \rho\log_2 \snr + \rho \log_2\frac{\alpha g}{\rho \LG} \qquad \mbox{(high SNR),}
\end{equation}
with $\alpha g$ the solution of \eqref{eq:AGCpoly}.  Note that the second term in the right-hand side of \eqref{eq:Casymptotic} depends only on $\rho$ and $\LG$, but not on the SNR. The pre-log factor in \eqref{eq:Casymptotic} is the ratio of the signal bandwidth to the system bandwidth $\rho \in \left[\frac{1}{2},1\right]$, and is not affected by self-interference, whose effect is a shift of the spectral efficiency vs. SNR curves. 



\section{Receiver Structures}
\label{sec:receiver}

In the general PD relay operation, its time-varying nature results in ICI in the frequency domain as seen in \eqref{eq:HX}, or equivalently \eqref{eq:Frelation2}.  The decoding of symbols transmitted through channels of the form \eqref{eq:Frelation2} is a well-studied problem, especially in MIMO systems \cite{tse2005}. 
As an alternative to the optimal maximum likelihood (ML) receiver, which achieves the spectral efficiency \eqref{eq:speff2} at the expense of potentially large computational complexity, in this section we analyze the performance of four suboptimal detectors: (i) the direct detection scheme which treats interference as noise, (ii) the zero-forcing (ZF) receiver, (iii) the Linear Minimum Mean Squared Error (LMMSE) receiver, and (iv) the Successive Interference Cancellation (SIC) receiver.

\subsection{Direct decoding}
\label{sec:direct}

The channel model \eqref{eq:Frelation2} can be rewritten as 
\begin{equation}\label{eq:sin}
\bar{\R}[q] = \sqrt{g}\D_N \bar{\X}[q] + \sqrt{g}(\T_N - \I_N)\D_N \bar{\X}[q] + \bar{\W}[q].
\end{equation}
The first and second terms in the right-hand side of \eqref{eq:sin} respectively represent the signal part (since the matrix $\sqrt{g}\D_N$ is diagonal) and the intercarrier interference (since  the matrix $\sqrt{g}(\T_N - \I_N)\D_N$ has zeros on the diagonal). The signal covariance matrix is $\frac{g \bar P_x}{\rho}\I_N$, whereas that of the interference plus noise is
$\C_{\rm I+N} \triangleq \frac{g \bar P_x}{\rho}\tilde{\T}_N\tilde{\T}^H_N + BN_0 \I_N$, where $\tilde{\T}_N \triangleq \T_N - \I_N$. Therefore, the achievable rate of a direct decoding strategy in which subcarriers are independently decoded with the interference term regarded as noise is given by
\begin{equation}
R_{\rm Direct} = \Delta f \sum_{n=1}^N \log_2\left(1+\rho_n^{\rm Direct}\right),
\end{equation}
where $\rho_n^{\rm Direct}$ is the signal-plus-interference-to-noise ratio (SINR) at the $n$-th subcarrier:
\begin{equation}\label{eq:rhondirect}
\rho_n^{\rm Direct} = \frac{g \bar P_x/\rho}{[\C_{\rm I+N}]_{n,n}} = \frac{\mu}{1+\mu\left[\tilde{\T}_N\tilde{\T}_N^H\right]_{n,n}},
\end{equation}
with $\mu$ as in \eqref{eq:mudef}. From \eqref{eq:Tentries}, the diagonal elements of $\tilde{\T}_N\tilde{\T}_N^H$ can be readily found: writing $n=(k-1)P+m$ with $k \in\{1,\ldots,K+1\}$ and $m\in\{1,\ldots,P\}$, 
\begin{equation}\label{eq:TTdiag}
\left[\tilde{\T}_N\tilde{\T}_N^H\right]_{n,n} = \alpha g \frac{1-(\alpha g)^{k-1}}{1-\alpha g},
\end{equation}
where $\alpha g$ is obtained from \eqref{eq:AGCpoly}.
Using \eqref{eq:rhondirect}-\eqref{eq:TTdiag}, and with $\delta(\rho)$ as in \eqref{eq:deltarho}, the corresponding spectral efficiency can be written as
\begin{eqnarray}
\frac{R_{\rm Direct}}{B} &=& \frac{1}{L}   \sum_{n=1}^N \log_2 \left(1+\frac{\mu}{1+\mu\left[\tilde{\T}_N\tilde{\T}_N^H\right]_{n,n}}\right) \\
&=& (1-\rho)\left[ \sum_{k=1}^K \log_2\left( 1+ \frac{\mu}{1+  \alpha g\frac{1-(\alpha g)^{k-1}}{1-\alpha g} \mu}\right) \right. \nonumber\\
& & \left.{}+ (1-\delta(\rho)) \log_2\left( 1+ \frac{\mu}{1+  \alpha g\frac{1-(\alpha g)^{K}}{1-\alpha g} \mu}\right) \right].
\label{eq:speff_direct}
\end{eqnarray}
In the absence of self-interference ($\alpha=0$), all log terms in \eqref{eq:speff_direct} are equal; and since $(1-\rho)(K+1-\delta(\rho)) = \rho$, \eqref{eq:speff_direct} becomes $ \rho \log_2\left(1+\frac{\snr}{\rho}\right)$, i.e., direct decoding is of course optimal, see \eqref{eq:speff_no_si}. On the other hand, when $\alpha > 0$, all of the log terms in \eqref{eq:speff_direct} tend to finite values as the SNR increases, except for the one corresponding to $k=1$ in the summation. Hence, in the high SNR regime, \eqref{eq:speff_direct} behaves as $(1-\rho) \log_2\snr + c$ with $c$ independent of $\snr$, so that the pre-log factor is now $1-\rho$. When compared with \eqref{eq:Casymptotic}, this shows the detrimental effect of self-interference when its structure is not exploited in the decoding process. As $\rho$ is increased, self-interference becomes more pronounced due to the larger overlap of the spectral supports of the relay input and output signals.

\subsection{ZF Receiver}
\label{sec:ZF}
In view of Lemma \ref{lem:inverse}, the inverse of the channel matrix in model \eqref{eq:Frelation2} is 
$(\sqrt{g}\T_N\D_N)^{-1} = \frac{1}{\sqrt{g}}\D_N^H (\I_N - \sqrt{\alpha g}e^{-j\theta_0}\S_N)$, which can be implemented with very low complexity. After application of this ZF receiver, each subcarrier is independently decoded. The achievable rate with the ZF receiver  is given by 
\begin{equation}
R_{\rm ZF} = \Delta f \sum_{n=1}^N \log_2\left(1+\rho_n^{\rm ZF}\right),
\end{equation}
where $\rho_n^{\rm ZF}$ is the signal-to-noise ratio at the $n$-th subcarrier. Since the covariance matrix of the post-processing noise is $\frac{BN_0}{g} \D_N^H (\T_N^H\T_N)^{-1}\D_N$, one has
\begin{equation}
\rho_n^{\rm ZF} = \frac{\bar P_x/\rho}{\frac{BN_0}{g} \left[\D_N^H (\T_N^H\T_N)^{-1}\D_N\right]_{n,n}}
= \frac{\mu}{\left[(\T_N^H\T_N)^{-1}\right]_{n,n}}.
\end{equation}
Using Lemma \ref{lem:inverse},  $(\T_N^H\T_N)^{-1} =  \I_N - \sqrt{\alpha g}e^{-j\theta_0}\S_N - \sqrt{\alpha g}e^{j\theta_0}\S_N^H +\alpha g \S_N\S_N^H$. The matrix $\S_N\S_N^H$ is diagonal, with the first $P=L-N$ diagonal elements equal to $0$ and the last $N-P$ equal to $1$. Therefore, $\left[(\T_N^H\T_N)^{-1}\right]_{n,n}$ equals $1$ for $n=1,\ldots,P$ and $1+\alpha g$ for $n=P+1,\ldots,N$. This results in
\begin{equation}\label{eq:speff_zf}
\frac{R_{\rm ZF}}{B} = \rho \log_2(1+\mu) + (2\rho-1) \log_2\left(\frac{1+\alpha g +\mu}{1+\alpha g+(1+\alpha g)\mu}\right), 
\end{equation}
which scales as $\rho\log_2\snr$ for high SNR. The pre-log factor, $\rho$, is the same as that of the optimal ML receiver, see \eqref{eq:Casymptotic},  and in contrast with $1-\rho$ for the direct decoding strategy of Sec.~\ref{sec:direct}. Note that the second term in the right-hand side of \eqref{eq:speff_zf} is non-positive (it is zero for $\alpha=0$) and constitutes a penalty term.

\subsection{LMMSE Receiver}
\label{sec:LMMSE}

Based on the channel model \eqref{eq:Frelation2}, the linear MMSE receiver computes $\hat{\X} = \F^H\bar{\R}$. The matrix $\F$ is chosen to minimize $\ex{ \lVert \bar{\X} - \hat{\X}\rVert^2}$, and is given by
\begin{equation}\label{eq:Fmmse}
\F = \frac{1}{\sqrt{g}}\left( \T_N\T_N^H + \mu^{-1} \I_N \right)^{-1} \T_N\D_N,
\end{equation}
with $\mu$ as in \eqref{eq:mudef}.
The achievable rate with the LMMSE receiver is given by
\begin{equation}
R_{\rm LMMSE} = \Delta f \sum_{n=1}^N \log_2\left(1+\rho_n^{\rm LMMSE}\right),
\end{equation}
with corresponding spectral efficiency
\begin{equation}\label{eq:speff_mmse}
\frac{C_{\rm LMMSE}}{B} = \frac{\rho}{N} \sum_{n=1}^N \log_2\left(1+\rho_n^{\rm LMMSE}\right), 
\end{equation}
where $\rho_n^{\rm  LMMSE}$ is the SINR at the $n$-th subcarrier. It is given by
\begin{equation}
\rho_n^{\rm  LMMSE} = \frac{q_n}{1-q_n},
\end{equation}
(see e.g. \cite{tse2005}), with $q_n$ the $(n,n)$ entry of the effective channel matrix $\F^H(\sqrt{g}\T_N\D_N)$:
\begin{eqnarray}
q _n &=& \left[\T_N^H \left( \T_N\T_N^H + \mu^{-1} \I_N \right)^{-1} \T_N \right]_{n,n} \\
&=& 1- \left[(\I_N + \mu \T_N^H\T_N)^{-1}\right]_{n,n}, \label{eq:qn}
\end{eqnarray}
for which it does not seem possible to obtain a closed-form expression. In the high SNR regime ($\mu\to\infty$), the MMSE receiver \eqref{eq:Fmmse} approaches the ZF receiver $(\sqrt{g} \T_N\D_N)^{-1}$, and therefore the spectral efficiency \eqref{eq:speff_mmse} approaches \eqref{eq:speff_zf} asymptotically as $\snr\to\infty$.



\subsection{Successive Interference Cancellation Receiver}
The channel matrix $\sqrt{g}\T_N \D_N$ in \eqref{eq:Frelation2} is lower triangular, which makes application of successive decoding particularly attractive. 
The sequence of symbols in the first carrier $\bar{\R}_1$ is decoded and used to substract $\bar{\X}_1$ from the next affected carrier, whose index is $L-N$. The process would continue until all carriers are decoded without interference. This SIC scheme is still suboptimal, because the power of the interference terms is not exploited. Since the diagonal elements of the channel matrix $\sqrt{g}\T_N \D_N$ have all square magnitude $g$, and the covariance matrices of $\bar{\X}$ and $\bar{\W}$ are $\frac{\bar P_x}{\rho} \I_N$ and $BN_0 \I_N$ respectively,  the achievable rate of the SIC receiver is readily found to be
\begin{equation}
R_{\rm SIC} = B_u \log_2\left(1+\frac{g\bar P_x}{\rho B N_0}\right).
\end{equation}
The corresponding spectral efficiency is therefore
\begin{equation}\label{eq:speff_sic}
\frac{R_{\rm SIC}}{B} = \rho\log_2(1+\mu) = \rho \log_2\left(1+ \frac{\alpha g}{\LG} \frac{\snr}{\rho}\right),
\end{equation}
with $\mu$ as in \eqref{eq:mudef} and $\alpha g$ given by the positive solution of \eqref{eq:AGCpoly}. Comparing \eqref{eq:speff_sic} with \eqref{eq:speff_no_si}, it is seen that the SIC receiver suffers an effective SNR reduction with respect  to the ideal case with no self-interference given by the factor $\frac{\alpha g}{\LG}= \frac{g\bar P_x}{\bar P_y}$, which is the ratio of the power of the useful signal component, $g\bar P_x$, to the total relay transmit power $\bar P_y$; the rest of the power $\bar P_y-g\bar P_x$ corresponds to the self-interference, as seen in \eqref{eq:iorelation}. Also note that for high SNR, \eqref{eq:speff_sic} achieves \eqref{eq:Casymptotic}, so the SIC receiver is optimal in high SNR, as could be expected.

\section{HD versus FD operation}
\label{sec:HDvsFD}

A high amount of self-interference can be expected to favor HD over FD operation, since the additional bandwidth will not compensate for the degradation due to the interference. 
Motivated by  the practical importance of the HD and FD operation modes, we quantify the spectral efficiency in both cases as a function of the SNR and loop gain. On the one hand, the HD spectral efficiency is easily computed: 
\begin{equation}\label{eq:speff_HD}
\left.\frac{C}{B}\right|_{\rm HD} = \frac{1}{2}\log_2\left(1+ \frac{\bar P_y}{N_0 B/2}\right) = \frac{1}{2} \log_2\left(1+ 2\, \snr \right),
\end{equation}
which of course is independent of $\LG$. Observe that \eqref{eq:speff_HD} can also be obtained from \eqref{eq:speff2} by taking $\rho=\frac {1}{2}$, since in that case one has $K=0$, $\delta(\rho) = 0$, and $\mu = 2\,\snr$.

On the other hand, the following result provides the FD spectral efficiency (assuming optimal ML decoding) in closed form (see Appendix \ref{app:FDspeff} for the proof):

\begin{lemma}\label{lem:FDspeff}
As $\rho \to 1$, the spectral efficiency \eqref{eq:speff2} becomes
\begin{equation}
\label{eq:speff_FD}
\left.\frac{C}{B}\right|_{\rm FD} = \log_2\left(\frac{1+\snr+2\,\LG + \sqrt{(1+\snr)^2 + 4\,\LG \,\snr}}{2(1+\LG)}\right).
\end{equation}
\end{lemma}

In the absence of self-interference ($\LG=0$), the FD spectral efficiency \eqref{eq:speff_FD} reduces to $\log_2(1+\snr)$, which is always larger than \eqref{eq:speff_HD} and approaches $2\cdot\left.\frac{C}{B}\right|_{\rm HD}$ for large $\snr$. On the other hand, as the loop gain increases ($\LG \to \infty$), the FD spectral efficiency falls to zero, showing the detrimental effect of self-interference in FD operation.

Consider now the direct decoding strategy (all self-interference is regarded as noise) applied to the FD case.
Using an approach analogous to that in Appendix \ref{app:FDspeff}, it can be shown that \eqref{eq:speff_direct} becomes
\begin{equation}\label{eq:speff_directFDasym}
\lim_{\rho\to 1} \frac{R_{\rm Direct}}{B} = \log_2\left(1 + \frac{\snr}{1+\LG+\LG\,\snr}\right),
\end{equation}
which is in agreement with the corresponding expression in \cite{Riihonen2009a}. Note that, as $\snr\to\infty$, \eqref{eq:speff_directFDasym} saturates at $\log_2\left(1+\frac{1}{\LG}\right)$.  

An important issue regarding the design of FD relays is that of transmit power optimization \cite{Riihonen2009a,Riihonen2011a}. If we regard $\bar P_y$ as the maximum available power at the relay and allow for power control, so that the relay transmit power is $P_y = p\,\bar P_y$ with $0 \leq p \leq 1$, then the spectral efficiency of the FD relay with direct decoding is given by \eqref{eq:speff_direct} upon substituting $\snr$ and $\LG$ by $p\,\snr$ and $p\,\LG$, respectively. It can be readily checked that the maximum of \eqref{eq:speff_directFDasym} is attained for $P_y = \min\left\{ \sqrt{\frac{BN_0\,P_x}{\alpha}}, \bar P_y\right\}$, yielding
\begin{equation}\label{eq:speff_directFDasymPC}
\lim_{\rho\to 1} \frac{R_{\rm Direct}}{B} = \left\{ \begin{array}{cc}
\log_2\left(1 + \frac{\snr}{1+\LG+\LG\,\snr}\right), & \mbox{if $\LG \leq \frac{1}{\snr}$}, \\
\log_2\left(1 + \frac{\snr}{\LG+2\sqrt{\LG\,\snr}}\right), & \mbox{if $\LG \geq \frac{1}{\snr}$}. 
\end{array}\right.
\end{equation}
Therefore, having the FD relay transmit at full power is not necessarily optimal when self-interference is regarded as noise, as observed in \cite{Riihonen2009a,Riihonen2011a,Leonardo2014a}: if the self-interference is sufficiently large (as determined by the condition $\LG \geq \frac{1}{\snr}$), then it is better to reduce the transmit power.
On the other hand, the expression resulting from replacing $\snr$ and $\LG$ by $p\,\snr$ and $p\,\LG$, respectively, in \eqref{eq:speff_FD} turns out to be monotonically increasing in $p\in [0,1]$; therefore, with ML decoding, having the relay transmit at full power is optimal. 

\begin{figure*}[!htb]
\centering
\includegraphics[width=0.6\textwidth]{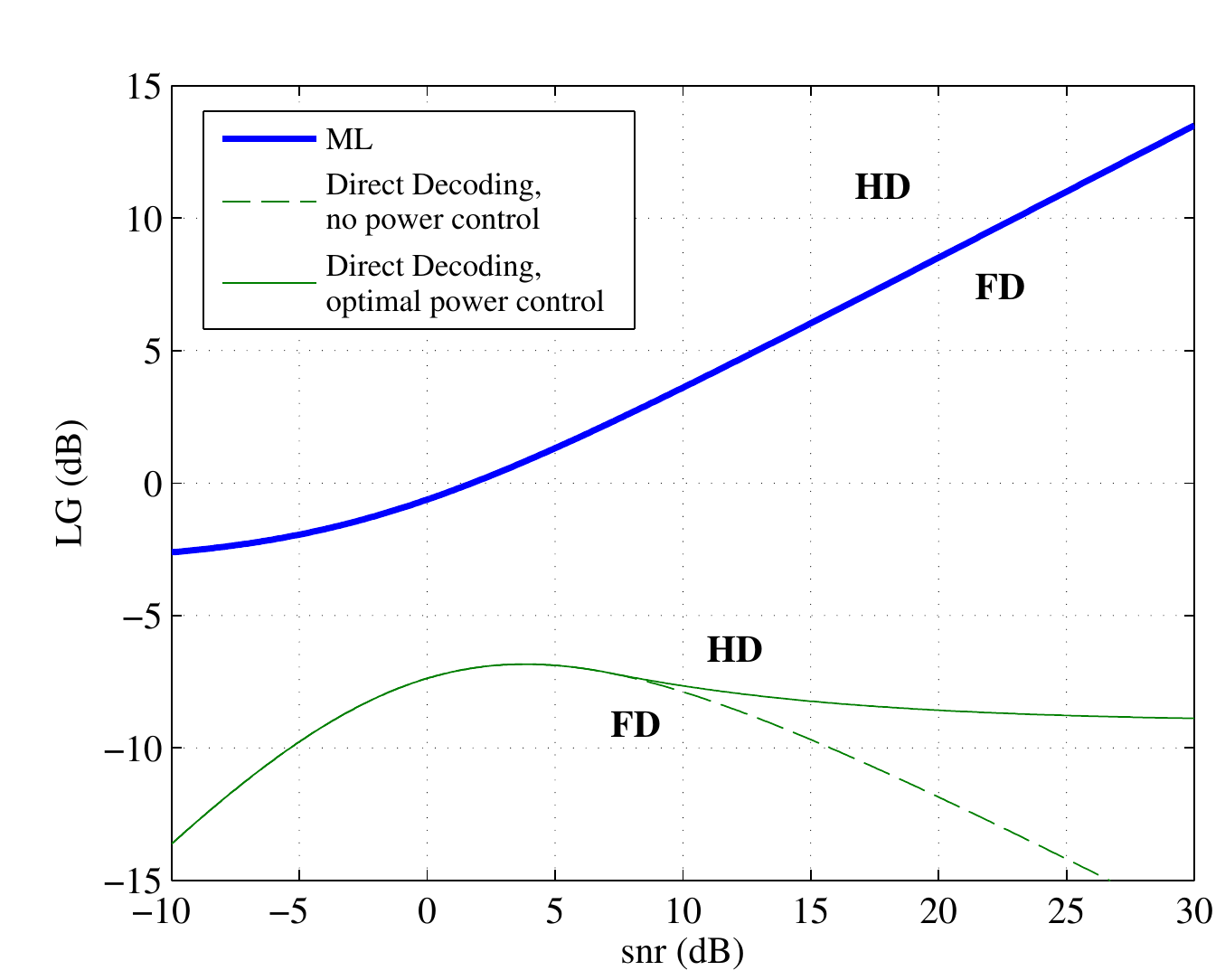}
\caption{HD vs. FD performance. For the ML and direct decoding strategies, the corresponding lines represent the boundary of the regions in which one of these modes outperforms the other.}
\label{fig:CompareFDvsHD}
\end{figure*}

The regions for which one operational mode (HD or FD) outperforms the other are depicted in Fig.~\ref{fig:CompareFDvsHD}, where the lines correspond to the boundary (set of points such that the spectral efficiencies of the FD and HD modes become equal) for ML and direct decoding (with and without power control for the latter).

\begin{itemize}
\item For ML decoding, and with asymptotically small SNR, this boundary is the horizontal line $\LG = \frac{1}{2}$, whereas the asymptote for large $\snr$ is $\LG = \sqrt{\frac{\snr}{2}}$. Note that whenever the loop gain is below $-3$ dB, FD outperforms HD regardless of the $\snr$ value. For loop gain values larger than $-3$ dB, the spectral efficiency of FD is larger for sufficiently high SNR. 

\item For direct decoding, the situation is quite different. Even with optimal power control, FD cannot perform better than HD as soon as $\LG> -7$ dB, regardless of the SNR. 

\end{itemize}


\section{Results}\label{sec:simulations}

\begin{figure*}[!p]
\centering
\begin{subfigure}{.49\textwidth}
\begin{centering}
\includegraphics[width=\textwidth]{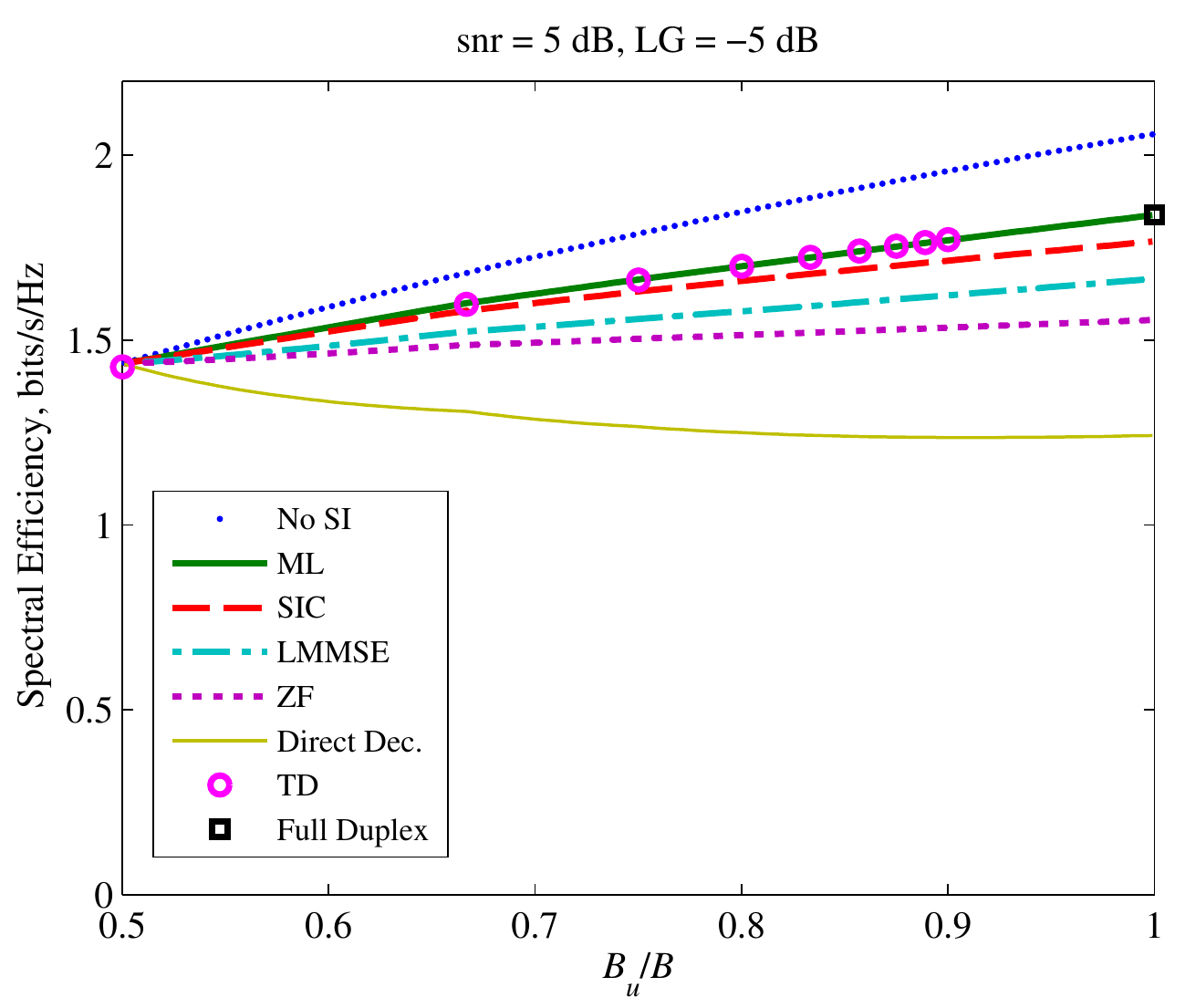}
\end{centering}
\end{subfigure}
\begin{subfigure}{.49\textwidth}
\begin{centering}
\includegraphics[width=\textwidth]{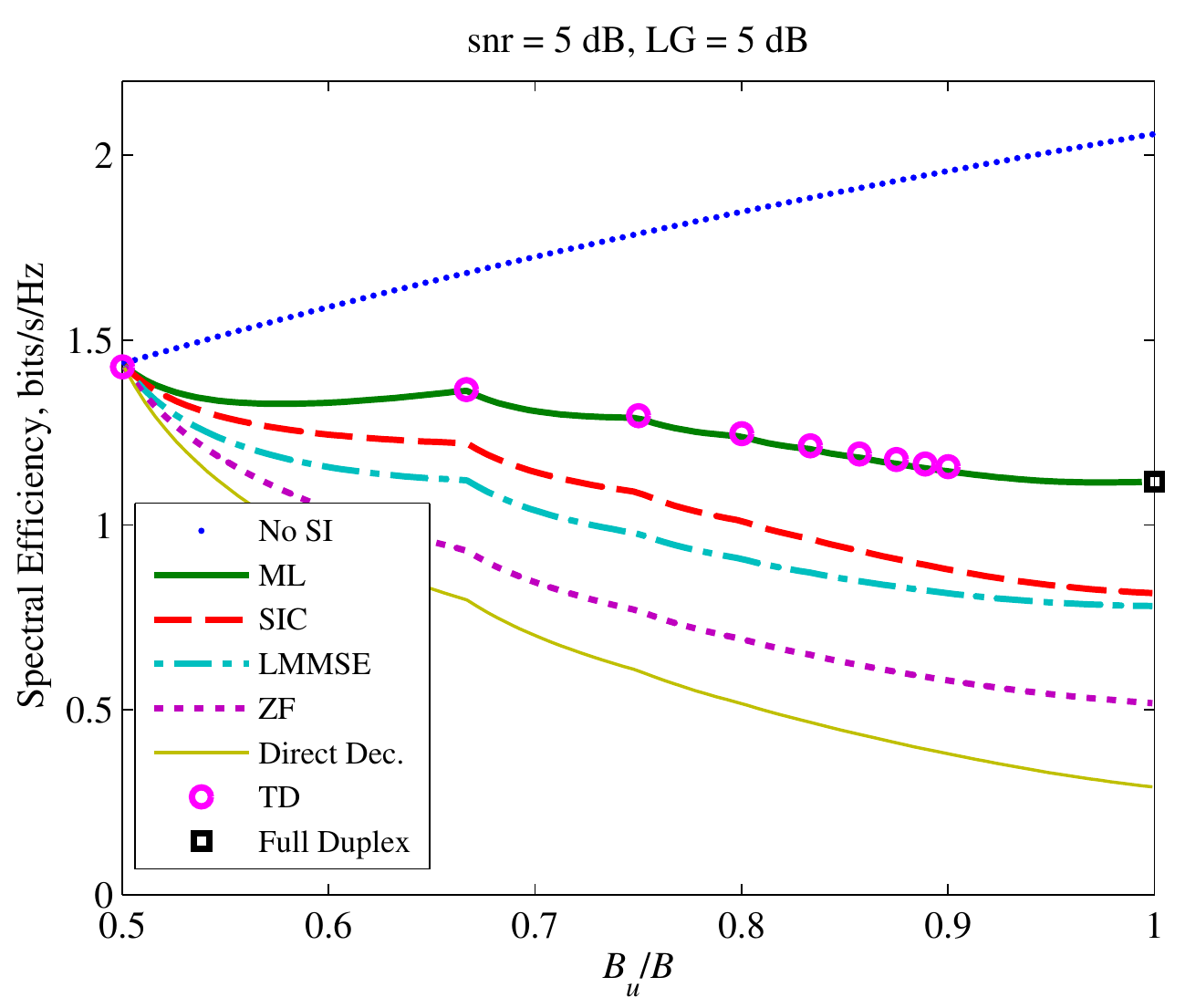}
\end{centering}
\end{subfigure}\\
\vspace*{-0.5cm}
\begin{subfigure}{.49\textwidth}
\begin{centering}
\includegraphics[width=\textwidth]{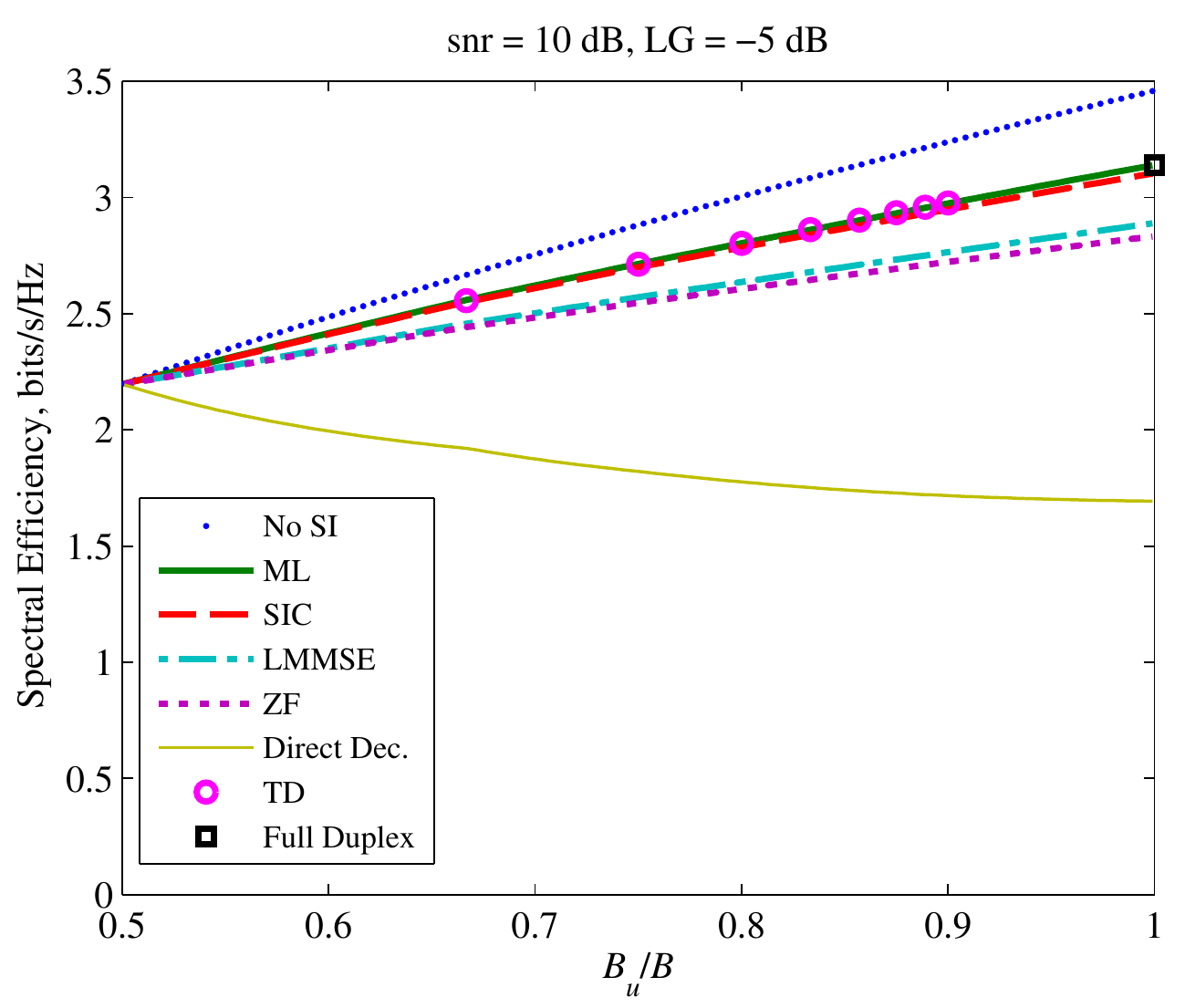}
\end{centering}
\end{subfigure}
\begin{subfigure}{.49\textwidth}
\begin{centering}
\includegraphics[width=\textwidth]{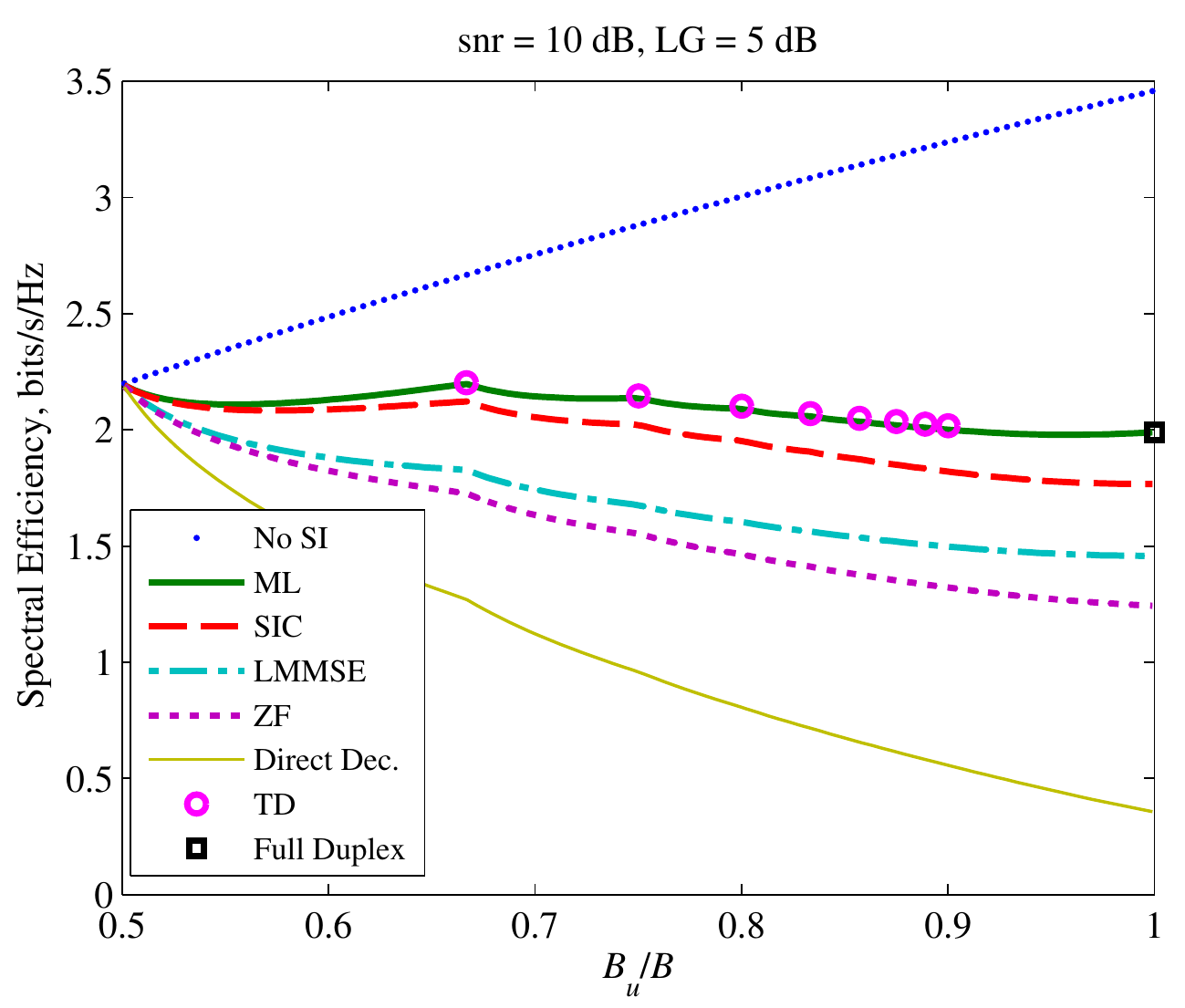}
\end{centering}
\end{subfigure}\\
\vspace*{-0.5cm}
\begin{subfigure}{.49\textwidth}
\begin{centering}
\includegraphics[width=\textwidth]{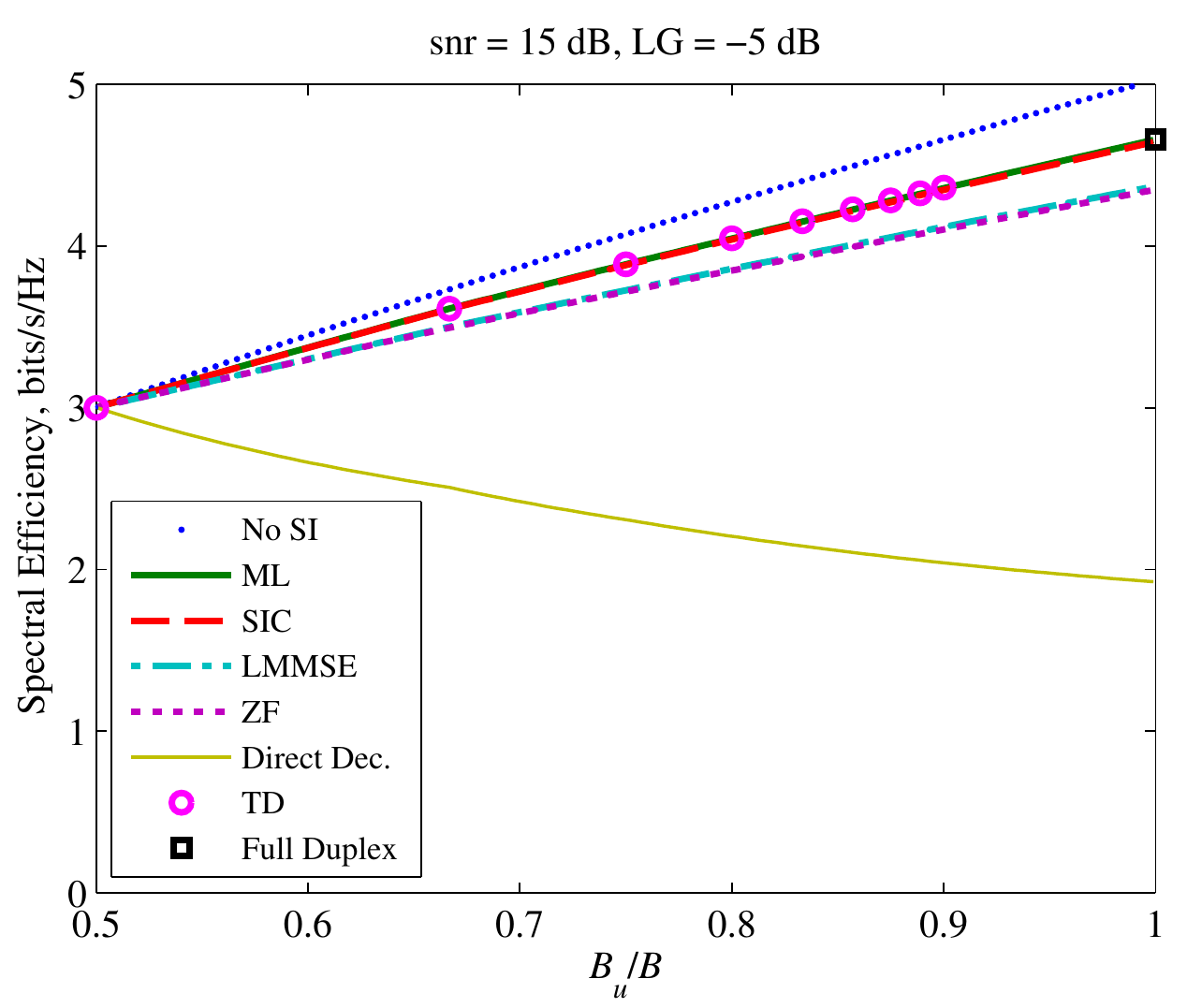}
\end{centering}
\end{subfigure}
\begin{subfigure}{.49\textwidth}
\begin{centering}
\includegraphics[width=\textwidth]{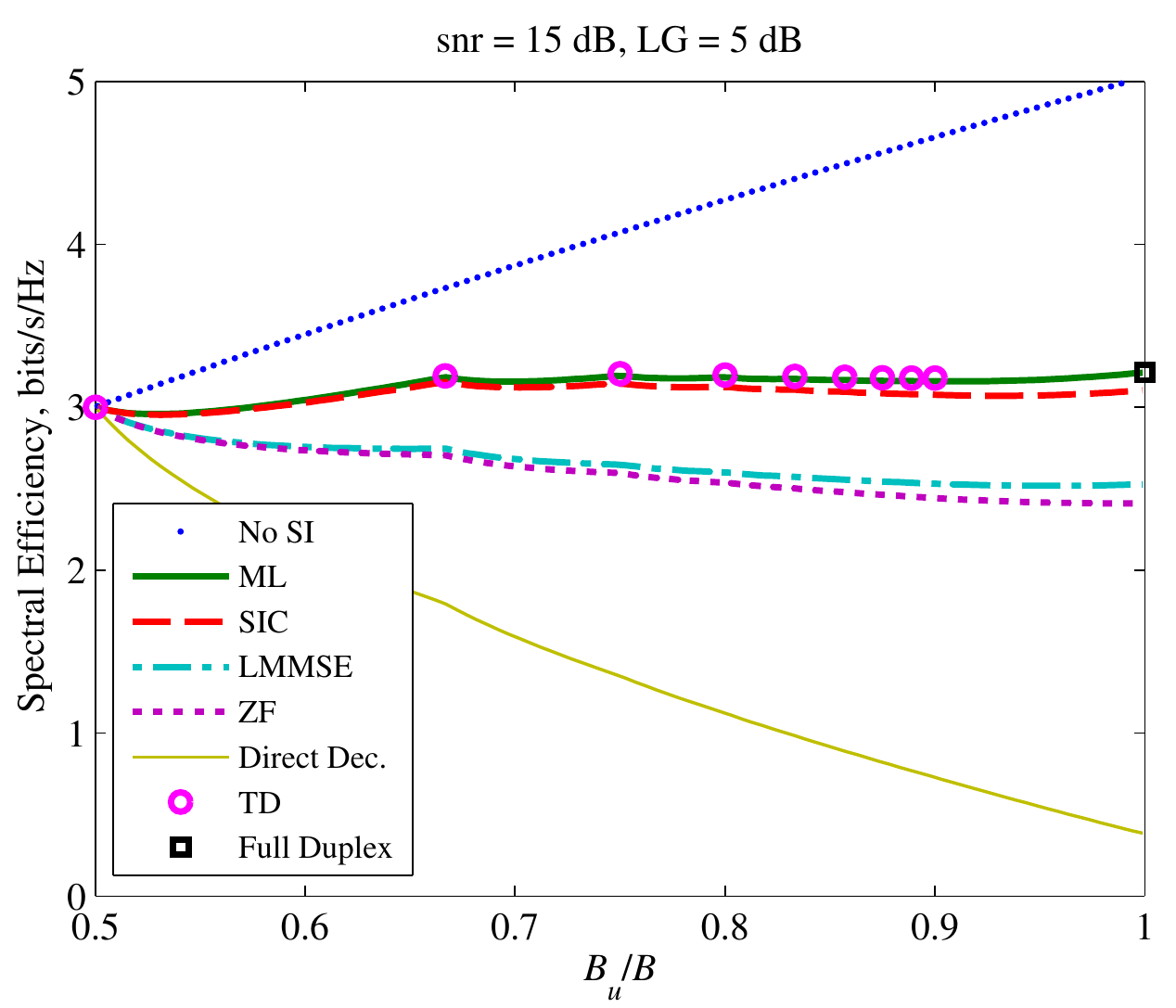}
\end{centering}
\end{subfigure}
\caption{Spectral efficiency of the A\&F PD relay vs. $\rho = B_u/B$. Left: weak coupling ($\LG=-5$ dB); right: strong coupling ($\LG = 5$ dB); $\snr = 5$ dB (top), $10$ dB (middle) and $15$ dB (bottom).}
\label{fig:SE_vs_rho}
\end{figure*}

Fig.~\ref{fig:SE_vs_rho} shows the spectral efficiency of the PD A\&F relay network with uniform power allocation across the input bandwidth $B_u$, and for different receiver strategies. The operation point is determined by the bandwidth ratio $\rho = B_u/B$, the signal-to-noise ratio $\snr = \bar P_y/(BN_0)$, and the magnitude of the coupling (loop gain) $\LG$. The different curves in Fig.~\ref{fig:SE_vs_rho} are as follows:
\begin{itemize}
\item {\it No SI:} this performance upper bound corresponds to the case without self-interference ($\alpha=0$), and is given by \eqref{eq:speff_no_si}.
\item {\it ML:} spectral efficiency of an optimum ML decoder, given by \eqref{eq:speff2}, and based on the frequency-domain approach of Sec.~\ref{sec:FDapproach}.
\item {\it SIC:} spectral efficiency with a SIC-based receiver, given by \eqref{eq:speff_sic}.
\item {\it LMMSE:} spectral efficiency with a linear MMSE-based receiver, computing as in Sec.~\ref{sec:LMMSE}. Since a closed-form expression is lacking, a total of $N=1000$ subcarriers was used in the numerical computations.
\item {\it ZF:} spectral efficiency with a ZF-based linear receiver, given by \eqref{eq:speff_zf}.
\item {\it Direct Dec.:} spectral efficiency with a direct decoding strategy, given by \eqref{eq:speff_direct}.
\item {\it TD:} These points are computed following the time-domain approach of Sec.~\ref{sec:TDapproach}, for values  of $\rho = \frac{N_{ch}-1}{N_{ch}}$, $N_{ch}=2,\ldots,10$. A truncated sinc was assumed for the filter $L_u(f)$, with delay $t_0$ satisfying $Bt_0 = 5N_{ch}$. The block size $M$ was taken as $M=\left(\left\lceil \frac{\ell_p+1}{N_{ch}}\right\rceil + \kappa\right) N_{ch}$, with $\kappa = 300$.
\item {\it Full Duplex:} spectral efficiency of the Full-Duplex relay, given by \eqref{eq:speff_FD}.
\end{itemize}

Fig.~\ref{fig:SE_vs_rho} shows results for three SNR values (5, 10 and 15 dB) and two coupling scenarios: $\LG = -5$ dB (weak coupling, i.e., self-interference power lower than signal power)  and $\LG = 5$ dB (strong coupling, i.e., self-interference power higher than signal power).

A close match is observed between the spectral efficiency values obtained via the time-domain approach of Sec.~\ref{sec:TDapproach} and those obtained under the frequency-domain approximation. Note that the latter approach does not rely on the existence of a periodic behavior of the PD relay and is computationally much simpler.

Regardless of the receiver strategy, and as soon as $\rho > \frac{1}{2}$, the performance of the PD relay significantly degrades in the presence of self-interference, as can be seen by comparing the left and right columns in Fig.~\ref{fig:SE_vs_rho}. This degradation is particularly severe for the direct decoding approach, which brings out the need to take self-interference into account at the receiver even in weak coupling scenarios. SIC detection provides close to optimal performance, except in low SNR conditions, as expected; and it outperforms LMMSE and ZF in general. 

The jagged appearance of the spectral efficiency curves with the bandwidth ratio $\rho$ in strong coupling scenarios is due to our assumptions of ideal brickwall filter responses and rectangular power spectral densities, which make the analysis tractable. As a result, the number of self-interference terms $K$ in \eqref{eq:Kmax} increases by one 
as $\rho$ crosses the values of the form $\frac{2}{3}$, $\frac{3}{4}$, $\frac{4}{5}$,\ldots, 
producing abrupt changes in the derivative of the spectral efficiency at those points. 

\begin{figure*}[!htb]
\centering
\includegraphics[width=0.8\textwidth]{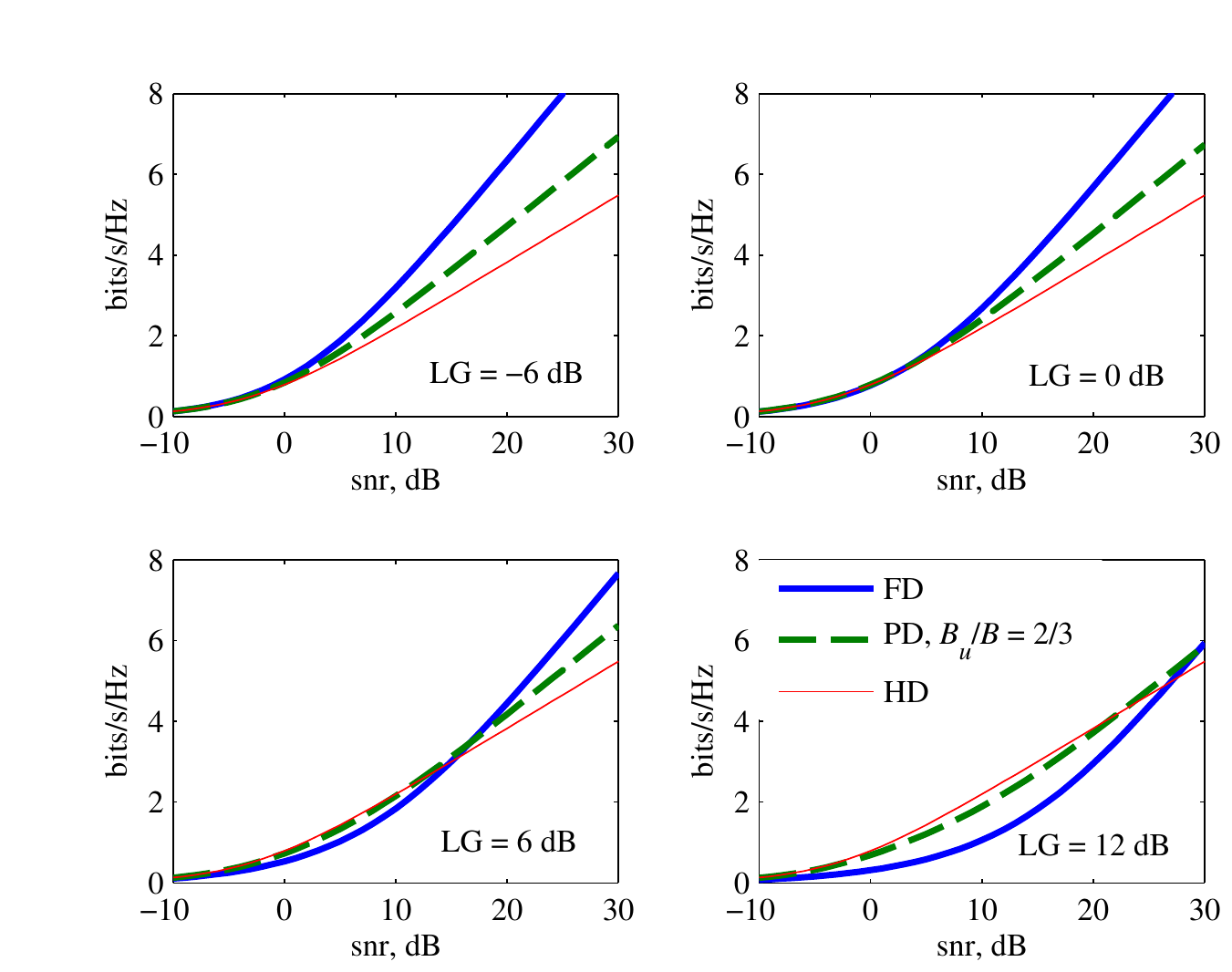}
\caption{Spectral efficiencies (for an optimum ML decoder) of the FD, PD ($\rho = \frac{2}{3}$) and HD modes vs. $\snr$.}
\label{fig:SEvsSNR}
\end{figure*}

\begin{figure*}[!htb]
\centering
\includegraphics[width=0.8\textwidth]{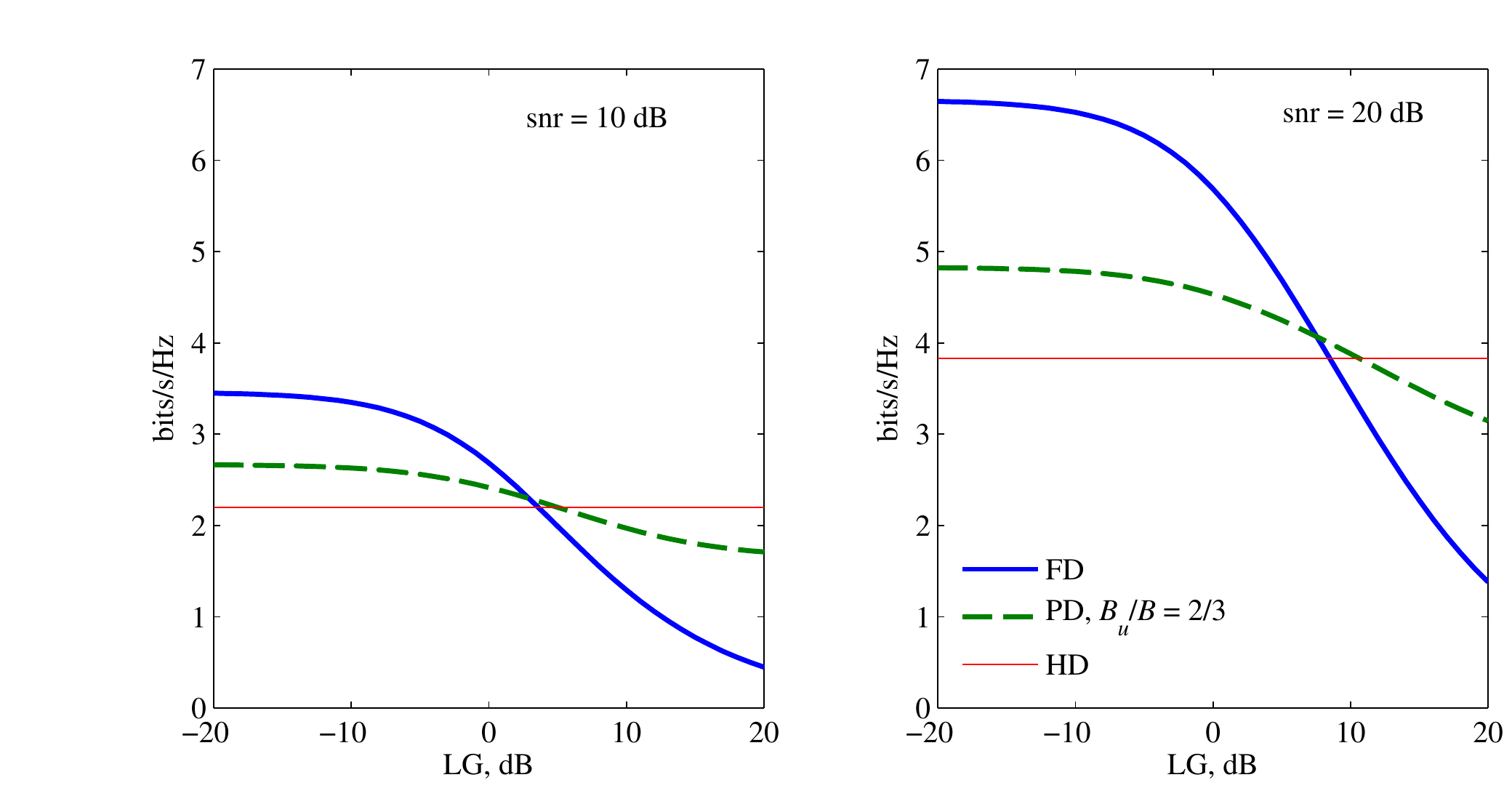}
\caption{Spectral efficiencies (for an optimum ML decoder) of the FD, PD ($\rho = \frac{2}{3}$) and HD modes vs. $\LG$.}
\label{fig:SEvsLG}
\end{figure*}

Under strong coupling, it is observed that the advantage of PD ($\rho >\frac{1}{2}$) with respect to HD ($\rho =\frac{1}{2}$) is in general small, if any. On the other hand, in weak coupling scenarios, spectral efficiency monotonically improves with the bandwidth ratio $\rho$. This improvement is obtained at the cost of the additional complexity required at the receiver to manage the (weak) self-interference, because the number $K$ of self-interference terms to handle increases with $\rho$, up to $K=\infty$ for $\rho = 1$ (FD mode). In this way, selection of an intermediate PD mode with $\frac{1}{2} < \rho < 1$ allows to trade off performance (in terms of spectral efficiency) and decoding complexity. As an example, in the setting of Fig.~\ref{fig:SE_vs_rho} with $\snr = 15$ dB and $\LG=-5$ dB, the PD modes corresponding to $\rho = \frac{2}{3}$ (for which $K=1$) and $\rho = \frac{3}{4}$ ($K=2$) provide improvements of up to 20\% and 30\% with respect to the HD mode, respectively. This is further illustrated in Figs.~\ref{fig:SEvsSNR} and~\ref{fig:SEvsLG}: even for $\LG=0$, i.e., a self-interference component of the same power as the signal, the PD mode with $\rho = \frac{2}{3}$ provides a sizable improvement with respect to HD if the SNR is sufficiently high, and with moderate receiver complexity relative to FD.

\section{Conclusions}\label{sec:conclusions}

An Amplify-and-Forward relay with partial overlap between input and output spectra, generalizing the well-known half-duplex and full-duplex cases, has been proposed and analyzed in the presence of self-interference at the relay.  
Its spectral efficiency for different overlap ratios has been obtained by exploiting the LPTV nature of this Partial Duplex  relay. The fact that this type of relay belongs to the class of bandwidth-preserving LPTV filters allowed to apply a frequency-domain approach which shows a remarkable accuracy to predict the true spectral efficiency. An important conclusion is that proper management of self-interference (which should not be simply regarded as noise) becomes mandatory in order to reap the benefits of spectrum overlap. With this in mind, several suboptimal decoding strategies at the receiver were also analyzed, among which Succesive Interference Cancellation emerges as a promising technique, performing close to the optimal ML decoder in high SNR. By effectively limiting the number of self-interference terms or "echoes", the proposed Partial Duplex mode provides a means to trade off spectral efficiency and receiver complexity.



\appendices

\section{Proof of Lemma \ref{lem:uniqueness}}
\label{app:uniqueness}
Assume $\alpha>0$. If $\rho=1$, then $K=+\infty$ and \eqref{eq:AGCpoly} reads as $\sum_{k=1}^\infty (\alpha g)^k = \LG$. This implies $0\leq\alpha g < 1$ (note that $\alpha g$ is the squared magnitude of the pole of the transfer function of the FD relay, which therefore is stable), and  $\frac{\alpha g}{1- \alpha g} = \LG$, from which $\alpha g = \frac{\LG}{1+\LG}$.

If $\rho<1$, consider the polynomial $p(s) = -\rho·\LG + \sum_{k=1}^{K+1} (1-k(1-\rho)) s^k$. First, note that the coefficients $1-k(1-\rho)$ are positive: for $k=1,\ldots, K+1$,
\begin{equation}\label{eq:coeffs1}
1-k(1-\rho) \geq 1-(K+1)(1-\rho) = 1- \left\lceil\frac{\rho}{1-\rho}\right\rceil (1-\rho).
\end{equation}
Write now $\left\lceil\frac{\rho}{1-\rho}\right\rceil = \frac{\rho}{1-\rho} + \delta$, with $\delta\in [0,1)$. Then \eqref{eq:coeffs1} reads 
\begin{equation}\label{eq:coeffs2}
1-k(1-\rho) \geq 1- (\rho + \delta(1-\rho)) = (1-\rho)(1-\delta) > 0,
\end{equation}
since $\rho < 1$. From the positivity of this terms it follows that $\lim_{s\to\infty} p(s) = +\infty$. 
Since $p(0) = -\rho\LG \leq 0$, $p(s)$ has at least a root in $s\in[0,+\infty)$. Moreover, 
$p'(s) =  \sum_{k=1}^{K+1} (1-k(1-\rho)) k s^{k-1} \geq 0$ for all $s\geq 0$, so that $p(s)$ is monotonically increasing in $s\in [0,+\infty)$. This implies that the root is unique.

Finally, if $\alpha=0$ (no self-interference), then $\LG=0$ as well, so \eqref{eq:AGCpoly} ceases to be informative. However, from \eqref{eq:traceTTH}, 
$\frac{1}{N}\trace\{\T_N\T_N^H\} = 1$, and substituting this in \eqref{eq:Pyg} yields $\bar P_y = g\bar P_x$.

\section{Proof of Theorem \ref{th:detQ}}
\label{app:detQ}
Given $P\in \mathbb{N}$, $a,b,\phi \in \mathbb{R}$ and $c \in \mathbb{C}$, let us define the matrices $\A_{M}(a,c,\phi) \in \mathbb{C}^{M\times M}$ and $\C_{M}(c,\phi) \in \mathbb{C}^{M\times P}$ as follows: for $M>P$,
\begin{equation}
\A_M(a,c,\phi) = \left(\begin{array}{cccccc} 
a & {\bf 0}_{P-1}^T & c e^{j\phi} & & & \\
{\bf 0}_{P-1} & \ddots & {\bf 0}_{P-1} & \ddots & & \\
c^*e^{-j\phi} & {\bf 0}_{P-1}^T & a & & \ddots & \\
 & \ddots & & \ddots & & c e^{j(M-P)\phi} \\
 &  & \ddots & & \ddots & {\bf 0}_{P-1} \\
 &  &  & c^*e^{-j(M-P)\phi} & {\bf 0}_{P-1}^T & a \end{array}\right),
\end{equation}
\begin{equation}
\C_M(c,\phi) = \left(\begin{array}{ccc} 
 & & \\ & {\bf 0}_{(M-P)\times P} &  \\ & & \\ \hline  ce^{j(M-P)\phi} & & \\ & \ddots & \\ & & c e^{jM\phi} 
\end{array}\right),
\end{equation}
whereas for $M\leq P$, $\A_M(a,c,\phi) = a\I_M$ and
\begin{equation}
\C_M(c,\phi) = \left(\begin{array}{ccc|ccc} 
 & & & ce^{j\phi} & & \\
 & {\bf 0}_{M\times (P-M)} & & & \ddots & \\
 & & & & & ce^{jM\phi}
\end{array}\right).
\end{equation}
From these, and for $N>P$, define now $\Z_N(a,b,c,\phi) \in \mathbb{C}^{N\times N}$ as
\begin{equation}\label{eq:defZ}
\Z_N(a,b,c,\phi) = \left(\begin{array}{cc} \A_{N-P}(a,c,\phi) & \C_{N-P}(c,\phi) \\ \C_{N-P}^H(c,\phi) & b\I_P 
\end{array}\right).
\end{equation}
Note that the matrix $\Q_N-\lambda\I_N$, with $\Q_N$ defined by \eqref{eq:Qa}-\eqref{eq:Qc}, can be written as $\Q_N - \lambda\I_N = \Z_N(a,b,c,\phi)$ with $a=1+\alpha g -\lambda$, $b=1-\lambda$, $c=-\sqrt{\alpha g}e^{j\theta_0}$ and $\phi= \phi_0$.

In order to compute $\det\Z_N(a,b,c,\phi)$, we use the expression for the determinant of partitioned matrices to obtain
\begin{equation}\label{eq:det_step1}
\det\Z_N(a,b,c,\phi) = b^P \det\left(\A_{N-P}(a,c,\phi) - \frac{1}{b}\C_{N-P}(c,\phi)\C_{N-P}^H(c,\phi)\right).
\end{equation}
If $P<N\leq 2P$, \eqref{eq:det_step1} readily evaluates to
\begin{equation}\label{eq:detZ1}
\det\Z_N(a,b,c,\phi) = b^P \left(a-\frac{|c|^2}{b}\right)^{N-P}, \qquad P<N\leq 2P.
\end{equation}
On the other hand, for $N>2P$, since
\begin{equation}
\C_{N-P}(c,\phi)\C_{N-P}^H(c,\phi) = \left(\begin{array}{cc}{\bf 0}_{(N-2P)\times (N-2P)} & \\ & |c|^2 \I_P \end{array}\right),
\end{equation}
it follows that 
\begin{equation}
\A_{N-P}(a,c,\phi) - \frac{1}{b}\C_{N-P}(c,\phi)\C_{N-P}^H(c,\phi) = \Z_{N-P}\left(a,a-\frac{|c|^2}{b},c,\phi\right),
\end{equation}
so that the following recursive relation is obtained:
\begin{equation}\label{eq:detZ2}
\det\Z_N(a,b,c,\phi) = b^P \det \Z_{N-P}\left(a,a-\frac{|c|^2}{b},c,\phi\right), \qquad N>2P.
\end{equation}
Let us define the scalar sequence
\begin{equation}
\tilde\gamma_1(a,b,c) = b, \qquad \tilde\gamma_n(a,b,c) = a - \frac{|c|^2}{\tilde\gamma_{n-1}(a,b,c)}, \quad n \geq 2.
\end{equation}
From \eqref{eq:detZ1} and \eqref{eq:detZ2}, one finds that, with $K=\left\lceil \frac{N}{P} \right\rceil -1$,
\begin{equation}\label{eq:detZtildegamma}
\det\Z_N (a,b,c,\phi) = \tilde\gamma_{K+1}^{N-KP}(a,b,c)\prod_{n=1}^{K}\tilde\gamma_n^P(a,b,c),
\end{equation}
which is independent of the phase angles $\phi$ and $\angle c$. Alternatively, let $\gamma_k(a,b,c) \triangleq \prod_{n=1}^k\tilde\gamma_n(a,b,c)$.  Then \eqref{eq:detZtildegamma} can be rewritten as
\begin{equation}\label{eq:detZgamma}
\det\Z_N (a,b,c,\phi) = \left[\gamma_{K}(a,b,c)\right]^{(K+1)P-N} \left[\gamma_{K+1}(a,b,c)\right]^{N-KP}.
\end{equation}
The terms $\gamma_k(a,b,c)$ can be obtained recursively as follows. One has
\begin{equation}
\gamma_1(a,b,c) = b, \qquad \gamma_2(a,b,c) = ab - |c|^2,
\end{equation}
whereas for $k >2$,
\begin{eqnarray}
\gamma_{k} = \gamma_{k-1} \tilde\gamma_{k} 
&=& \gamma_{k-1}  \left(a-\frac{|c|^2}{\tilde\gamma_{k-1}}\right)  \nonumber\\
&=&  \gamma_{k-1}  \left(a-\frac{|c|^2\gamma_{k-2}}{\gamma_{k-1}}\right)  \nonumber\\
&=&  a \gamma_{k-1} - |c|^2 \gamma_{k-2}. \label{eq:recursion_gamma}
\end{eqnarray}
Particularizing this recursion for $a=1+\alpha g -\lambda$, $b=1-\lambda$, $|c|^2=\alpha g$, the result in Theorem \ref{th:detQ} is proved.

\section{Proof of Lemma \ref{lem:FDspeff}}
\label{app:FDspeff}
In order to compute the asymptotic value of \eqref{eq:speff2} as $\rho \to 1$, we take $\rho = \frac{N}{N+1}$ and make $N\to\infty$. With this choice of $\rho$, one has $K+1=N$ and $\delta(\rho)=0$, so that, from \eqref{eq:speff2},
$\frac{C}{B} = \frac{1}{K+2} \log_2q_{K+1}(-\mu)$. Clearly, as $K\to\infty$, $q_{K+1}(-\mu)$ must diverge, or otherwise $\frac{C}{B}$ would go to zero. In fact, for $\frac{C}{B}$ to have a finite and positive limit as $K\to \infty$, one must have $q_{K}(-\mu) \approx a b^{K}$ asymptotically, for some finite $a>0$, $b>1$, and the spectral efficiency becomes
\begin{equation}
\lim_{K\to \infty} \frac{\log_2q_{K+1}(-\mu)}{K+2}  = \lim_{K\to\infty} \left(\frac{\log_2 a}{K+2}  + \frac{K+1}{K+2} \log_2 b \right) = \log_2b.
\end{equation}
To find $b$, note that $b= \lim_{k\to \infty} \frac{q_{k}(-\mu)}{q_{k-1}(-\mu)}$. From the recursion \eqref{eq:recursionq},  one has
\begin{equation}\label{eq:recursion_ratio}
\frac{q_k(-\mu)}{q_{k-1}(-\mu)} = (1+\alpha g + \mu) - \alpha g \frac{q_{k-2}(-\mu)}{q_{k-1}(-\mu)}.
\end{equation}
Now, from Lemma \ref{lem:uniqueness}, one has $\alpha g \to \frac{\LG}{1+\LG}$ for $\rho \to 1$. Using this and the definition \eqref{eq:mudef}, it follows that $\mu \to \frac{\snr}{1+\LG}$ as $\rho \to 1$. Therefore, taking the limit as $k\to \infty$ in both sides of \eqref{eq:recursion_ratio},
\begin{equation}\label{eq:recursion_ratio}
b = \left(1+\frac{\LG}{1+\LG}  + \frac{\snr}{1+\LG} \right) - \frac{\LG}{1+\LG} \frac{1}{b}.
\end{equation}
This quadratic equation has the following solutions:
\begin{equation}\label{eq:bsols}
b = \frac{1+\snr+2\,\LG \pm \sqrt{(1+\snr)^2 + 4\,\LG \,\snr}}{2(1+\LG)}.
\end{equation}
The solution corresponding to the minus sign in \eqref{eq:bsols} can be easily shown to be no larger than 1, whereas the one corresponding to the plus sign is no smaller than 1 (both are equal to one iff $\snr=0$). Therefore, the former can be discarded, and the latter yields \eqref{eq:speff_FD}.

\bibliographystyle{IEEEtran}
\bibliography{FullDuplex}

\end{document}